\PassOptionsToPackage{table,xcdraw,x11names}{xcolor}

\documentclass[acmsmall,screen]{acmart}

\usepackage{algorithmic}
\usepackage{graphicx}
\usepackage{xspace}
\usepackage{pifont} 
\usepackage{enumitem}
\usepackage[many]{tcolorbox}
\usepackage{multirow}
\usepackage{booktabs} 
\usepackage[normalem]{ulem}
\useunder{\uline}{\ul}{}
\usepackage{threeparttable} 
\usepackage{tikz}
\usepackage{float}
\usepackage{framed}
\usepackage{quoting}
\usepackage[justification=centering]{caption}

 \colorlet{shadecolor}{LavenderBlush2}
\usepackage{lipsum}
\newenvironment{shadedquotation}
 {\begin{shaded*}
  \quoting[leftmargin=0pt, vskip=0pt]
 }
 {\endquoting
 \end{shaded*}
}

\AtBeginDocument{%
  }

\usepackage{microtype}
\newcommand{\rs}{RT\xspace}
\newcommand{\Rs}{Remediation Tactics\xspace}
\newcommand{\RS}{Remediation Tactics\xspace}
\newcommand{\pv}{\textit{Maintainer}\xspace}
\newcommand{\sk}{\textit{User}\xspace}
\newcommand{\va}{\textbf{AdjVer}\xspace}
\newcommand{\patch}{\textbf{Patch}\xspace}
\newcommand{\altlib}{\textbf{AltLib}\xspace}
\newcommand{\bypass}{\textbf{Bypass}\xspace}
\newcommand{\chgconfig}{\textbf{ChgConfig}\xspace}
\newcommand{\reinit}{\textbf{Re-init}\xspace}
\newcommand{\nosolu}{\textbf{NoSolu}\xspace}
\definecolor{OliveGreen}{rgb}{0,0.6,0} 
\newcommand{\ly}[1]{\textcolor{black}{#1}}
\newcommand{\rev}[1]{\textcolor{black}{#1}}

\newcommand*\emptycirc[1][1ex]{\tikz\draw (0,0) circle (#1);} 
\newcommand*\halfcirc[1][1ex]{%
  \begin{tikzpicture}
  \draw[fill] (0,0)-- (90:#1) arc (90:270:#1) -- cycle ;
  \draw (0,0) circle (#1);
  \end{tikzpicture}}

\DeclareRobustCommand{\mybox}[2][gray!20]{%
\begin{tcolorbox}[
        breakable,
        left=0pt,
        right=0pt,
        top=0pt,
        bottom=0pt,
        colback=#1,
        colframe=#1,
        width=\linewidth, 
        enlarge left by=0mm,
        boxsep=5pt,
        arc=0pt,outer arc=0pt,
        ]
        #2
\end{tcolorbox}
}

\usepackage{makecell}
\usepackage{threeparttable}
\makeatletter  
\newif\if@restonecol  
\renewcommand\footnoterule{%
	\kern-3\p@
	\hrule\@width\columnwidth
	\kern2.6\p@}

\setcopyright{acmlicensed}
\copyrightyear{2025}
\acmYear{2025}
\acmDOI{XXXXXXX.XXXXXXX}

\acmConference[ISSTA '25]{International Symposium on Software Testing and Analysis}{June 25--28,
  2025}{Trondheim, Norway}
\acmISBN{978-1-4503-XXXX-X/18/06}

\makeatletter
\let\@authorsaddresses\@empty
\makeatother

\begin{document}

\title{Fixing Outside the Box: Uncovering Tactics for Open-Source Security Issue Management
}

\author{Lyuye Zhang}
\authornote{Lyuye Zhang and Jiahui Wu have equal contributions.}
\affiliation{%
  \institution{Nanyang Technological University}
  \city{Singapore}
  \country{Singapore}
}
\orcid{0000-0003-3087-9645}
\email{zh0004ye@e.ntu.edu.sg}

\author{Jiahui Wu}
\authornotemark[1]
\affiliation{%
  \institution{Nanyang Technological University}
  \city{Singapore}
  \country{Singapore}
}
\orcid{xxxx}
\email{JIAHUI004@e.ntu.edu.sg}

\author{Chengwei Liu}
\authornote{Chengwei Liu is the corresponding author.}
\affiliation{%
  \institution{Nanyang Technological University}
  \city{Singapore}
  \country{Singapore}
}
\orcid{xxxx}
\email{chengwei.liu@ntu.edu.sg}

\author{Kaixuan Li}
\affiliation{%
  \institution{East China Normal University}
  \city{Shanghai}
  \country{China}
}
\orcid{0000-0002-3517-353X}
\email{kaixuanli@stu.ecnu.edu.cn}

\author{Xiaoyu Sun}
\affiliation{%
  \institution{The School of Computing, Australian National University}
  \city{Canberra}
  \country{Australia}
}
\orcid{0000-0001-7434-0452}
\email{Xiaoyu.Sun1@anu.edu.au}

\author{Lida Zhao}
\affiliation{%
  \institution{Nanyang Technological University}
  \city{Singapore}
  \country{Singapore}
}
\orcid{xxxx}
\email{LIDA001@e.ntu.edu.sg}

\author{Chong Wang}
\affiliation{%
  \institution{Nanyang Technological University}
  \city{Singapore}
  \country{Singapore}
}
\orcid{xxxx}
\email{chong.wang@ntu.edu.sg}
\author{Yang Liu}
\affiliation{%
  \institution{Nanyang Technological University}
  \city{Singapore}
  \country{Singapore}
}
\orcid{xxxx}
\email{yangliu@ntu.edu.sg}


\begin{abstract}
In the rapidly evolving landscape of software development, addressing security vulnerabilities in open-source software (OSS) has become critically important. 
However, existing research and tools from both academia and industry mainly relied on limited solutions, such as vulnerable version adjustment and adopting patches, to handle identified vulnerabilities. However, far more flexible and diverse countermeasures have been actively adopted in the open-source communities. A holistic empirical study is needed to explore the prevalence, distribution, preferences, and effectiveness of these diverse strategies. 

To this end, in this paper, we conduct a comprehensive study on the taxonomy of vulnerability remediation tactics (\rs) in OSS projects and investigate their pros and cons.
This study addresses this oversight by conducting a comprehensive empirical analysis of 21,187 issues from GitHub, aiming to understand the range and efficacy of remediation tactics within the OSS community. We developed a hierarchical taxonomy of 44 distinct \rs and evaluated their effectiveness and costs. Our findings highlight a significant reliance on community-driven strategies, like using alternative libraries and bypassing vulnerabilities, 44\% of which are currently unsupported by cutting-edge tools. Additionally, this research exposes the community's preferences for certain fixing approaches by analyzing their acceptance and the reasons for rejection. It also underscores a critical gap in modern vulnerability databases, where 54\% of CVEs lack fixing suggestions—a gap that can be significantly mitigated by leveraging the 93\% of actionable solutions provided through GitHub issues.
\end{abstract}




\maketitle

\section{Introduction}

\rev{With its flexibility and transparency, open-source software (OSS) has become an integral part of modern software development. However, its widespread adoption also introduces significant security challenges.
For example, in 2021, a critical vulnerability in the widely used Log4j library~\cite{log4jnews1, log4jnews2, log4jnews3} affected millions of devices worldwide, underscoring the severe impact vulnerabilities can have. Such incidents highlight the urgent need for effective remediation strategies.
In the context of OSS security, remediation encompasses a series of actions aimed at mitigating risks posed by vulnerabilities~\cite{wu2023understanding, wang2023plumber, zhang2023mitigating}. This process is essential for both OSS developers and downstream users~\cite{bandara2020fix, bandara2021large, pan2024unveil, panichella2021won, piantadosi2019fixing}. Specifically, remediation~\cite{remediationpedia, bandara2021large} involves providing fixes for reported vulnerabilities. By addressing vulnerabilities, remediation enhances OSS security, ensures stability for downstream users, and helps developers maintain the reliability of the software supply chain.}

To achieve better remediation, researchers from both academia and industry have put years of effort into investigating and improving vulnerability remediation for software projects.
For instance, some researchers focus on the lifecycles of vulnerabilities in OSS projects to reveal the effectiveness and timeliness of vulnerability remediation~\cite{pan2024unveil, forootani2022exploratory, bandara2021large, bandara2020fix}. Other researchers concentrate on the prioritization of vulnerabilities for remediation due to concerns on cost-efficiency~\cite{panichella2021won, anvik2006should, shah2022vulnerability}. Moreover, the specific vulnerability remediation solutions, such as vulnerable version adjustment~\cite{liu2022demystifying, zhang2023compatible, zhang2023mitigating}, vulnerability patching~\cite{li2017large, 10.1145/3196884}, and workarounds~\cite{huang2016talos} are also well-studied by existing research. However, existing studies focus primarily on mainstream remediation tactics in academia, overlooking the diverse solutions employed within the OSS community and leaving numerous approaches to security issue remediation unexplored.


\rev{Regarding remediation tools (automated tools that process reported vulnerabilities and generate fixing suggestions), numerous tools and advisories have been developed to support security issue remediation. These tools are often integrated into Software Composition Analysis (SCA) tools, vulnerability analyzers, and vulnerability databases.}
Eclipse Steady~\cite{steady} and Coral~\cite{zhang2023compatible} aid developers in identifying and addressing vulnerabilities in open-source libraries by upgrading or downgrading libraries to their secure versions. 
For industry tools, OSV-fix~\cite{osvfix} and Snyk~\cite{snyk} streamline the process of fixing vulnerabilities in OSS dependencies directly in users' Software-Bill-of-Material (SBOM) files by modifying versions. 
SonarQube and Semgrep also incorporate remediation suggestions, such as effort estimation and plausible patches, for identified vulnerabilities in user projects. 
Vulnerability databases, such as OSV~\cite{OSV} and GitHub Advisory~\cite{githubadvisory}, may also provide remediation suggestions for reference for a limited number of vulnerabilities. Nevertheless, as demonstrated by our study (i.e., Section~\ref{sec: tool study} and~\ref{sec:rq4}), existing solutions are still largely limited to well-known and straightforward solutions, such that they either heavily rely on upgrading the affected versions and applying patches or rarely offers remediation suggestions. In comparison with the countermeasures (as unveiled by our preliminary study in Section~\ref{sec: tool study}) from OSS communities, there is a clear shortage of solutions incorporated in existing tools.

\rev{Based on the above findings and observations of existing industrial solutions and academic research, it is evident that current approaches heavily depend on established data sources, such as NVD, and primarily focus on mainstream tactics, such as upgrading libraries and patching source code. Additionally, as discussed in Section~\ref{sec:advisory_rs}, $54\%$ of Common Vulnerabilities and Exposures (CVEs) do not provide remediation suggestions. Given the limited remediation solutions available to millions of downstream users of vulnerable libraries, the adoption and effectiveness of existing fixes may be insufficient to ensure the security of the software community.
Fortunately, remediation solutions emerging from the community offer a much broader spectrum and can serve as a valuable complement to traditional approaches, enhancing both coverage and effectiveness in addressing security vulnerabilities. 
For example, issues~\cite{githubexample1,githubexample2} in Section~\ref{sec:party} and~\ref{sec:reasonsrejection} provide upgrading the library and environment respectively to fix the CVEs (CVE-2023-44487~\cite{CVE-2023-44487} and CVE-2024-31479~\cite{CVE-2024-31479}) that lack official remediation suggestions, demonstrating how community-driven solutions can effectively complement existing vulnerability databases.
}

To this end, we aim to conduct a comprehensive study to first investigate the existing remediation solutions, and then identify unrevealed possible countermeasures in the OSS community. Lastly, we explore whether they are adaptable to a wider range of vulnerabilities to enhance existing vulnerability-fixing tactics. 
Specifically, we perform an empirical study to effectively bridge this gap by analyzing 21,187 security issues on GitHub and constructing a hierarchical \RS (\rs) taxonomy that encompasses 44 distinct operations. Based on the taxonomy, we iteratively explored the community feedback and associated costs of \rs as well as the distinctions between \rs from GitHub and vulnerability databases to address the following research questions:

\noindent\textbf{RQ1: What are common remediation tactics for security issues in OSS projects?} It aims to categorize the remediation tactics and construct a taxonomy of them. 

\noindent\textbf{RQ2: How many remediation tactics were accepted or rejected and Why?} The distribution of accepted tactics was plotted and the reasons for the rejection were summarized.

\noindent\textbf{RQ3: How much do remediation tactics cost
regarding the time and manpower?} We measured the effectiveness and cost of \rs based on the timelines of the proposals.

\noindent\textbf{RQ4: What are the differences between remediation tactics provided by vulnerability databases and those from the community?} The \rs provided by vulnerability databases were compared with community solutions.

From the RQs, we found a significant portion of community-driven \rs (44\% of studied issues) remains unsupported by current tools. Key insights emphasize the need for \rs to be tailored to different roles—downstream users prefer simplicity, while maintainers value stability. 
Moreover, the absence of \rs in 54\% of CVEs in modern vulnerability databases underscores the urgent need for support for a broad spectrum of \rs. We made the following contributions:
\begin{itemize}[leftmargin=5pt]
    \item To the best of our knowledge, we conducted the first empirical study on the \rs of security issues in the community and constructed a hierarchical taxonomy. 
    \item We methodologically investigated the acceptance of \rs for different audiences of \rs. Then we categorized the reasons behind them regarding the associated costs.
    \item We evaluated the discrepancy between \rs in the community and vulnerability database advisories and revealed the potential complement among multiple sources.
    \item We provide the largest ever dataset of security issues in the academic world as well as the systematic analysis built on top of it, available at our repository~\cite{dataset}.
    
\end{itemize}



\section{Related Works}

\subsection{Vulnerability Fixing for OSS}
Fixing vulnerabilities in OSS has been considered the last mile for vulnerability management~\cite{remediation}.
Many research works have been conducted to investigate the fixing tactics of OSS vulnerabilities from various aspects.
Specifically, some researchers focused on the lifecycles of vulnerabilities existing in OSS projects to reveal the timeliness of vulnerability remediation. 
Pan et al.~\cite{pan2024unveil} distinguished the lifespans of critical and non-critical software vulnerabilities and found that critical vulnerabilities receive higher priority.
Forootani et al.~\cite{forootani2022exploratory} explored the timeline of self-fixed vulnerabilities in OSS projects and found that self-fixed vulnerabilities are resolved more quickly than non-self-fixed ones.
Bandara et al.~\cite{bandara2021large, bandara2020fix}, from the perspective of vulnerability correlation, identified that 18\% of vulnerability fixing commits actually introduced new vulnerabilities.
Moreover, 
some researchers also focus on the criteria and prioritization of vulnerabilities to optimize remediation tactics for projects. 
Piantadosi et al.~\cite{anvik2006should} compiled statistics on practitioners, timelines, and commits involved in remediation in two Apache projects.
Shah et al.~\cite{shah2022vulnerability} studied the vulnerability selection criteria for remediation based on Multiple Attribute Value Optimization.

Moreover, many researchers also focus on detailed countermeasures and effects of vulnerability remediation. 
First, as the most adopted solution, \textbf{adjusting vulnerable versions} has been studied by many researchers. 
Liu et al.~\cite{liu2022demystifying} proposed DTReme to exhaustively explore all possible dependency resolutions to find possible version-adjustment-based remediation tactics for NPM projects, 
Zhang et al.~\cite{zhang2023compatible} joint major concerns, such as compatibility~\cite{zhang2022has}, vulnerability reachability~\cite{wu2023understanding}, and dependency conflicts~\cite{wang2021will}, to propose an integrated solution for vulnerability remediation for Java projects. 
They also inspected the phenomenon of vulnerability persistence in most Java projects and proposed Ranger to remediate vulnerability impact from the ecosystem perspective. 
\textbf{OSS Patching} is also widely studied, 
Li et al.~\cite{li2017large} investigated the timeline for developing patches and found that security patches typically have a smaller footprint in code bases compared to non-security bug patches.
Farris et al.~\cite{10.1145/3196884} introduced the VULCON, which takes various metadata information of vulnerabilities to prioritize patching.
\textbf{Workaround}~\cite{khazaei2018vuwadb}, considered a special form of remediation, also provides temporary and non-ideal fixes for vulnerabilities. Huang et al.~\cite{huang2016talos} proposed Talos to generate workarounds for rapid vulnerability response. 

\subsection{Remediation Tactics in SOTA Tools}
\label{sec: tool study}
\rev{We explored three types of security tools that may provide vulnerability-fixing solutions based on the granularity of the code where the vulnerability resides. Specifically, three granularity levels are considered: code snippet, library, and whole program, each corresponding to the following types of tools: \ding{172} Automatic Vulnerability Repair (AVR) tools; \ding{173} Software Composition Analysis (SCA) tools; \ding{174} Vulnerability analysis (VA) tools; In our comparison, we have incorporated both popular academic and commercial tools to the best of our knowledge.}
\rev{To evaluate their remediation capabilities for comparison, we aimed to summarize three attributes: suggestions, fixing targets, and the requirements for providing the remediation. Specifically, we manually reviewed the documentation of them and tested them with provided sample cases, where available, where we ensured that the summarized attributes were at least supported by the tools' documentation or publication. The process was individually cross-checked by the first three authors, with the final summarized attributes determined using a majority voting scheme.}

Most SCA tools~\cite{wu2023ossfp,woo2021centris} provide follow-up \rs for detected vulnerabilities, thus included for further analysis~\cite{zhao2023software,hu2024empirical}. As listed in recent work~\cite{li2023comparison,esposito2024extensive,zhang2023open,abdulghaffar2023enhancing,sun2024llm4vuln}, famous vulnerability analysis tools, such as SonarQube~\cite{sonarqube}, Semgrep~\cite{semgrep}, Infer~\cite{infer}, SpotBugs~\cite{spotbugs}, and CodeQL~\cite{codql}, ZAP~\cite{owaspzap}, mostly analyze and report vulnerabilities in target programs with only SonarQube and Semgrep have remediation-related suggestions.
Popular AVR tools, such as VulRepair~\cite{fu2022vulrepair}, VulMaster~\cite{zhou2024out}, VulnFix~\cite{zhang2022program}, FootPatch~\cite{van2018static}, and ACFix~\cite{zhang2024acfix} usually fix vulnerabilities by generating patches. 

A variety of SOTA tools are listed in Table~\ref{tab:tools}. For remediation suggestions, SOTA tools mostly adopt established solutions like upgrading and patching. The vulnerabilities that these tools focus on are mostly CVEs as well as other self-collected security issues, such as GitHub Advisory Database (GHSA). 
SCA tools generally set predefined requirements for smooth remediation with compatibility as the primary one.
Beyond CVEs, numerous other security issues are disclosed within the OSS community. These discussions frequently introduce a variety of remediation solutions neglected by existing tools, highlighting different concerns and requirements. Our study seeks to bridge this gap by conducting a comprehensive analysis that incorporates firsthand discussions of a broad spectrum of security issues from the community.
\begin{table}[]
\setlength{\tabcolsep}{6pt}
\scriptsize
\caption{\rev{Comparison of SOTA Vulnerability Remediation Tools}}
\label{tab:tools}
\begin{tabular}{@{}llll|llll@{}}
\toprule
\rowcolor[HTML]{EFEFEF} 
{\color[HTML]{000000} \textbf{VA}}  & {\color[HTML]{000000} \textbf{Suggestion}} & {\color[HTML]{000000} \textbf{Target}} & {\color[HTML]{000000} \textbf{Req.}} & {\color[HTML]{000000} \textbf{SCA}} & {\color[HTML]{000000} \textbf{Suggestion}}  & {\color[HTML]{000000} \textbf{Target}} & {\color[HTML]{000000} \textbf{Req.}} \\ \midrule
\rowcolor[HTML]{FFFFFF} 
{\color[HTML]{000000} Semgrep~\cite{semgrep}}      & {\color[HTML]{000000} Patch}               & {\color[HTML]{000000} Vul}             & {\color[HTML]{000000} None}          & {\color[HTML]{000000} Steady~\cite{steady}}       & {\color[HTML]{000000} Upgrading}            & {\color[HTML]{000000} CVE}             & {\color[HTML]{000000} Compatibility} \\
\rowcolor[HTML]{EFEFEF} 
{\color[HTML]{000000} SonarQube~\cite{sonarqube}}    & {\color[HTML]{000000} N.A.}                & {\color[HTML]{000000} Vul}             & {\color[HTML]{000000} Time}          & {\color[HTML]{000000} Coral~\cite{zhang2023compatible}}        & {\color[HTML]{000000} Upgrading}            & {\color[HTML]{000000} CVE}             & {\color[HTML]{000000} Compatibility} \\
\rowcolor[HTML]{FFFFFF} 
{\color[HTML]{000000} Infer~\cite{infer}}        & {\color[HTML]{000000} N.A.}                & {\color[HTML]{000000} Vul}             & {\color[HTML]{000000} None}          & {\color[HTML]{000000} Dependabot~\cite{dependabot}}   & {\color[HTML]{000000} Upgrading}            & {\color[HTML]{000000} CVE, GHSA}       & {\color[HTML]{000000} Compatibility} \\
\rowcolor[HTML]{EFEFEF} 
{\color[HTML]{000000} SpotBugs~\cite{spotbugs}}     & {\color[HTML]{000000} N.A.}                & {\color[HTML]{000000} Vul}             & {\color[HTML]{000000} None}          & {\color[HTML]{000000} OSV fix~\cite{osvfix}}      & {\color[HTML]{000000} Upgrading}            & {\color[HTML]{000000} CVE, OSV}        & {\color[HTML]{000000} None}          \\
\rowcolor[HTML]{FFFFFF} 
{\color[HTML]{000000} CodeQL~\cite{codql}}       & {\color[HTML]{000000} Patch rule}          & {\color[HTML]{000000} Vul}             & {\color[HTML]{000000} SARIF}         & {\color[HTML]{000000} Snyk~\cite{snyk}}         & {\color[HTML]{000000} Upgrading, patch}     & {\color[HTML]{000000} CVE, SnykV}      & {\color[HTML]{000000} Compatibility} \\ \cmidrule(r){1-4}
\rowcolor[HTML]{EFEFEF} 
{\color[HTML]{000000} \textbf{AVR}} & {\color[HTML]{000000} \textbf{Suggestion}} & {\color[HTML]{000000} \textbf{Target}} & {\color[HTML]{000000} \textbf{Req.}} & {\color[HTML]{000000} Mend~\cite{mend}}         & {\color[HTML]{000000} Upgrading, patch}     & {\color[HTML]{000000} CVE}             & {\color[HTML]{000000} None}          \\ \cmidrule(r){1-4}
\rowcolor[HTML]{FFFFFF} 
{\color[HTML]{000000} VulRepair~\cite{fu2022vulrepair}}    & {\color[HTML]{000000} Patch}               & {\color[HTML]{000000} Vul}             & {\color[HTML]{000000} Syntax}        & {\color[HTML]{000000} Sonatype~\cite{sonatype}}     & {\color[HTML]{000000} Upgrading, migration} & {\color[HTML]{000000} CVE}             & {\color[HTML]{000000} Customized}    \\
\rowcolor[HTML]{EFEFEF} 
{\color[HTML]{000000} VulMaster~\cite{zhou2024out}}    & {\color[HTML]{000000} Patch}               & {\color[HTML]{000000} Vul}             & {\color[HTML]{000000} Syntax}        & {\color[HTML]{000000} OWASP~\cite{owasp}}        & {\color[HTML]{000000} Upgrading, Config}    & {\color[HTML]{000000} CVE}             & {\color[HTML]{000000} None}          \\
\rowcolor[HTML]{FFFFFF} 
{\color[HTML]{000000} VulFix~\cite{zhang2022program}}       & {\color[HTML]{000000} Patch}               & {\color[HTML]{000000} Vul}             & {\color[HTML]{000000} Syntax}        & {\color[HTML]{000000} Veracode~\cite{veracode}}     & {\color[HTML]{000000} Upgrading, patch}     & {\color[HTML]{000000} CVE, Debt}       & {\color[HTML]{000000} None}          \\ \bottomrule
\end{tabular}
\end{table}

\begin{figure*}[]
\centering
  \includegraphics[width=1\linewidth]{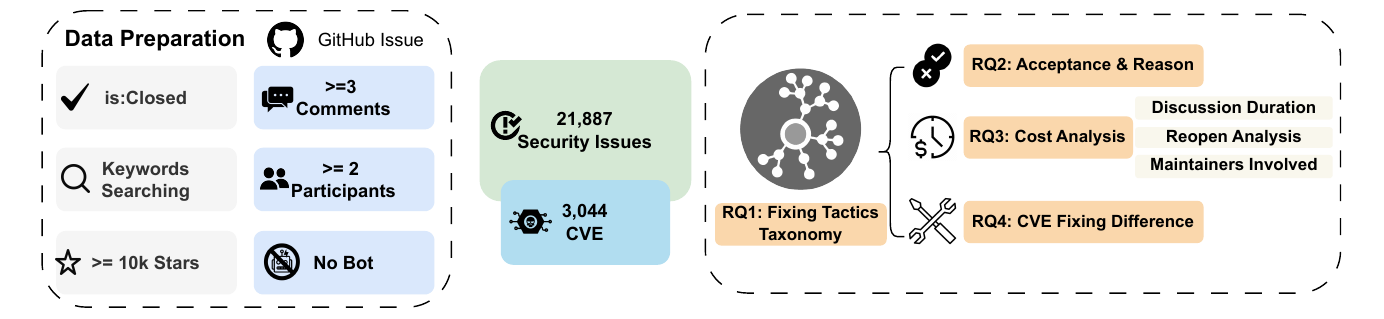}
  \caption{Overview of This Empirical Study}
  \label{fig:overview}
\end{figure*}

\section{Data Collection Methodology}
\label{sec:data}
\rev{Our focus is on security issues, as they pose significant risks to vulnerable libraries and downstream users. Given the widespread use of open-source software, popular projects (e.g., highly starred repositories) have a greater impact, making effective security fixes essential for safeguarding modern software systems. GitHub, as the largest source code hosting platform, hosts numerous popular OSS projects with significant developer involvement. As shown in Figure~\ref{fig:overview}, we selected GitHub as the primary source for our dataset to explore human-driven solutions to security issues, leveraging the wisdom-of-the-crowd approach beyond traditional \rs. We focused on the \textit{Issue} section for its flexible \rs proposals, while also referencing \textit{PRs} and \textit{commits} linked to issues to enrich the diversity of \rs in our study.}


GitHub Issue selection criteria have been systematically established by previous work, such as filtering out unresolved open issues~\cite{panichella2021won,zhang2023compatible}, filtering issues by community engagement (i.e., activeness)~\cite{jiang2017understanding, zhang2023compatible}, and excluding noisy information introduced by bots~\cite{wessel2022bots,wessel2021don}.
Honoring the criteria from previous work, we detail our selection of issues as follows: 

\noindent\rev{\ding{172} \textbf{Keywords Derivation}:} \rev{To ensure keywords are commonly used in security-related issues, they were obtained from the most frequent terms in patch commit messages of CVEs associated with GitHub patch links found in reference links. Since CVE patch commits on GitHub consistently aim to address security vulnerabilities and do not introduce platform-specific discrepancies, we utilized this commit message corpus for keyword extraction.
We first identified 8,782 CVEs published in 2024 and filtered out those without patch links, reducing the dataset to 645 CVEs. From these, we collected 854 patch links for commit message extraction. Finally, we ranked the top 100 most frequent terms from these commit messages.
(The lists of CVEs, URLs, and keywords are lemmatized and provided on our website), excluding irrelevant English stopwords and common words to focus on verbs and nouns, which yielded the selected keywords. Note that the term \textit{bug} was included because security issues are sometimes referred to as security bugs.}

\noindent\rev{\ding{173} \textbf{Setup searching parameters:}
To avoid including unresolved issues, we crawled \textit{closed issues} from GitHub dated from January 1, 2022, to May 31, 2024. As our focus is \rs regarding security issues, we selected the related keywords for searching via GitHub API, including combining two groups of words, \textit{security/vulnerability/issue/bug} and \textit{fix/remediation/tweak/repair/resolve} to encompass as many relevant issues as possible by depicting the nouns and verbs. These two groups were combined into a single query following the GitHub API documentation.}
\ly{\rev{
Since keyword searching on GitHub does not guarantee that all retrieved issues are related to our target security issues~\cite{liu2018adaptive}, we manually verified their relevance to avoid introducing excessive irrelevant cases by randomly sampling 500 from the returned issues. The majority voting among the first three authors was employed to resolve conflicted labels. We found that 86.6\% of issues were related to security fixes, while the rest involved bug fixes or performance tweaks. To preserve the integrity of the GitHub search results and avoid bias, we included all issues in the statistics as they were part of the \rs.
}}

\noindent\rev{
\ding{174} \textbf{Issue Filtering:} To guarantee that the issue has been actively discussed, conservatively at least 3 comments and 2 distinct 
participants should be involved. Issues with three comments are considered active because we assumed a complete issue-fixing process should at least include 3 conversations: proposal, verification, and acknowledgment.
We initially derived $293,934$ issues. Since our focus is on human-involved issues and human-developed \rs, issues created by bots were subsequently filtered out, resulting in $212,047$ issues (GitHub provides entries in API response indicating the bot status). This number is still beyond our capability. Thus we further narrow down to highly starred repositories as we assume highly-starred repositories typically exhibit more rigorous mechanisms of resolving issues. Ranking repositories in descending star counts, we established a cutoff at 2,600 manageable repositories with $21,187$ issues to align with our capacity for manual analysis.
}


\noindent\rev{\ding{175} \textbf{CVE-ID Labeling:} Within the issues from GitHub, those related to CVE were explicitly marked with CVE-ID tags. These CVE-related issues incorporated with the rest are used in RQ1,2,3. And only CVE-related issues are used in RQ4. Specifically, we employed two steps to pinpoint CVE-related issues. 
(1) We retrieved issues that mentioned at least one CVE-ID in their titles or comments. 
(2) Subsequently, the first two authors manually reviewed these issues to exclude issues not explicitly addressing the mentioned CVEs. The procedures were \ding{172} reading the sentences mentioning CVE-IDs to determine if the CVEs are fixing targets or simply referenced as related vulnerabilities. \ding{173} for the seemingly fixing targets CVEs, we approached NVD entries for them to cross-check if the vulnerabilities were discussed in the issues. If either condition was not satisfied, the issues were filtered out. 
Eventually, $2,919$ out of $4,462$ issues were collected. In total, $3,044$ CVE-IDs were identified and saved. \rev{Note that some issues (Example~\cite{multicveinanissue}) may be related to more than one CVE, as participants explicitly mentioned multiple CVEs being addressed within the discussion.}
}

\section{Empirical Study of Remediation in OSS}
In this section, we methodically explore and respond to the research questions through a series of iterative investigations, each yielding multiple insightful findings.

\subsection{RQ1: \RS}\label{sec:rq1}
In this RQ, the procedures to categorize the \rs from collected issues are elaborated, and the distribution of the summarized taxonomy is then demonstrated.

\subsubsection{Procedures of \RS Labeling}
\label{sec:categorization}
We conducted systematic labeling and review to summarize \rs based on Hybrid Card Sorting~\cite{hybridcardsorting}, a common approach for classification, and cross-review, involving four authors with over 5 years of experience in OSS security. 

\ly{\rev{\ding{172} Labeling: First, we randomly divided the issues into three folds, each containing 7,196 issues, and the first three authors conducted card sorting separately for each fold. No pre-defined cards were provided to encourage free ideation. Annotators reviewed issue titles and bodies, summarizing tactics into phrase-based cards and grouping them accordingly. Following the Hybrid Card Sorting approach, annotators could assign an issue to an existing card if appropriate. After individual labeling, 245 unique cards were generated. We then collaboratively de-duplicated them based on synonyms and similar concepts, ultimately distilling 44 cards which have been used to update all the labels if they are different.}}

\rev{The following quoted example demonstrates the clue we depended on to label the \rs. Because the creation comment mentions the new \texttt{Putty} has the vulnerability fixed and implies it could be upgraded to mitigate threats, the issue involves the version changing thus major category is \va. 
Then, because the version is upgraded based on the `just release` expression, indicating the newer version, the minor category is \textit{Upgrading to a secure version}.
Lastly, although \texttt{Putty} is the forked code base, it is used as a library in this project, the last category should be \textit{Upgrading vulnerable library to a secure version.}}

\begingroup
\begin{shadedquotation}
\footnotesize
\noindent\rev{
Quoting from an example Issue~\cite{githubexample1}}:\\
\rev{\textbf{Issue Creator}: Hi,
Just to bring up that Putty just released 0.81 which fixes the vulnerability detailed in the CVE article where a malicious server operator can, with enough connections, form the private key from the user.
Might want to rebuild against it ASAP.}

\noindent\rev{\textbf{Maintainer}: Yeah, I would really appreciate some testers. I just like to give it some time to bake before I officially call it released. Given the UI elements and hardware token interactions, I don't have a good automated testing pipeline... I depend on you all :-)}
\end{shadedquotation}
\endgroup
\ly{\rev{\ding{173} Cross-Review: 
At the beginning of labeling, we noticed that the labels were prone to mistakes due to the limited initial card set, which would expand over time due to the hybrid (open \& closed) nature of the process. This is a common challenge in open/hybrid card sorting~\cite{hybridcardsorting,hybridcardsortingchallenge1,hybridcardsortingchallenge2}. 
Therefore, four authors were involved in this label review to mitigate this challenge.
We conducted a cross-review along with majority voting for samples in each fold from the previous step. Note that the sorted cards from three folds in the Labeling step were de-duplicated and merged for the reference of reviewers in this step. Another annotator instead of the initial one was assigned as a reviewer to review the card selected for the issues. Given the sorted cards, if the reviewer agrees with the initial label, then the labeling is settled, otherwise, the reviewer refers to the fourth author as the judge, and the judge will gather labeling results from two annotators. After a majority vote among the three (initial annotator, reviewer, judge), the final label will be decided. In line with this, the first three authors independently reviewed three folds that were different from the ones they initially labeled. The fourth author was responsible solely for resolving conflicts during this process.
To balance efficiency and effectiveness, the cross-review was conducted on samples from each fold. Since the categorization largely converged after approximately 2,000 issues, we conservatively selected 3,000 random issues per fold for review. In total, 9,000 issues were reviewed, with 1,342 cases identified as inconsistent (14.91\% of the total). After inspecting the specific cases, we noticed that 1,301 of them (96.94\%) are actually consistently labeled on minor categories. Furthermore, the mistakes are predominantly concentrated in the initial issues. }}

\begin{figure}
\centering
  \includegraphics[width=0.9\linewidth]{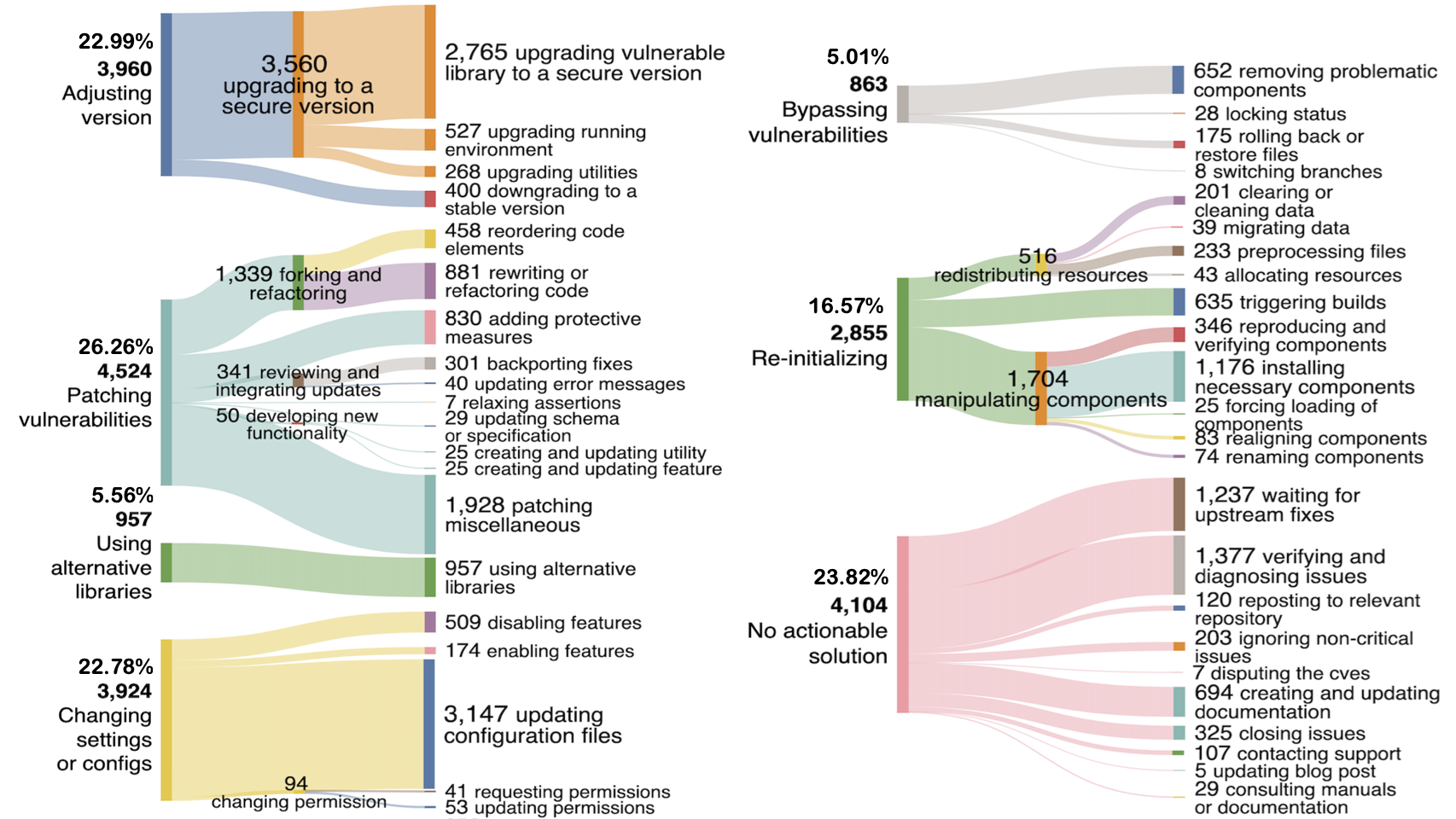}
  \caption{Taxonomy of \RS}
  \label{fig:taxo}
\end{figure}
\subsubsection{Analysis of \RS Taxonomy}
The initial Card Sorting produced a list of categories; however, overlapping and encompassment existed. To address this ambiguity, we developed a hierarchical taxonomy of \rs categories, as detailed in Figure~\ref{fig:taxo}. The taxonomy consists of three levels: the first level represents the major categories of tactics, the second level comprises subcategories within the first, and the third level includes specific operations commonly performed within these subcategories. Not every category contains subcategories or operations if there is no proper one. The shares of each category are visually represented by the widths of bars in Figure~\ref{fig:taxo}.
Besides \textit{No actionable solution} suggests the security issues cannot be fixed for now, the rest is actionable solutions that can be generally categorized into 6 major ones as follows: 
\begin{itemize}[leftmargin=3pt]
    \item \textbf{Version adjustment} (\va) usually refers to adjusting the versions of software components or dependencies used in the software projects by changing the versions in Bill-of-Material (BOM) files~\cite{bom}, such as Project Object Model (POM)~\cite{pom} for Maven~\cite{maven} and \texttt{package.json} for NPM~\cite{npm}.
    This category does not involve changes in source code files but only BOM files.
    As \textit{Version adjustment} only modifies the versions without removing the components, the subcategories include \textit{upgrade} and \textit{downgrade}. 
    
    \item \textbf{Patching vulnerabilities} (\patch) refers to patching the source code or binary of the software projects instead of external configurations. The GitHub issues in this major category mostly involve file changes by mentioning related commits or PRs in the issue events. Notably, this major category is the most frequently employed strategy, but the intentions of patches vary greatly. Due to the diversity of patches, we only roughly listed the commonly appeared subcategories.
    
    \item \textbf{Using alternative secure libraries} (\altlib) involves substituting an independent library or component within software projects. It is important to note that this category does not encompass the replacement of files or code snippets, which are classified under \textit{Patching vulnerabilities}. The component or library being replaced should have identifiable names, and the replacement process can be executed either through BOM files or by replacing a group of files. This category has a pretty small share as it demands the availability of alternative libraries and usually requires adaptation of the API invocation.
    
    \item \textbf{Changing settings or configurations} (\chgconfig) is not the same as \textit{Patching vulnerabilities} as it does not involve changing the source code. It commonly changes the external configuration files or configurations on the application before and at runtime. Since configurations or settings may be located in BOM files, we identified \chgconfig based on what has been changed in the BOM files. If the BOM files do not modify existing components, it is deemed as \chgconfig. Due to the diversity, this category has various subcategories and operations and the feasibility depends on the context of software projects.

    \item \rev{\textbf{Bypassing vulnerabilities} (\bypass) refers to running the program in a different way to avoid executing the vulnerable part. Since vulnerability bypassing is pure subtractive, it is considered an independent category. The pre-condition is that the vulnerable components are not in use or the bypass does not affect the program, which usually requires manual review and testing verification.}

    \item \textbf{Re-initializing} (\reinit) involves starting anew with fresh components and data arrangements. This category encompasses three common subcategories: components, data, and building. We labeled an issue as \reinit based on comments suggesting that the program should be restarted or refreshed, or words to that effect. Subcategories were then assigned by summarizing the preparatory steps mentioned in the comments prior to re-initialization.

    \item \textbf{No actionable solution} (\nosolu) does not belong to the major categories of \rs. It refers to suggestions that do not directly address the security issues immediately, such as \textit{Wait for upstream fixes} and \textit{Ignoring non-critical issues}. Since some of these suggestions have been accepted by the audience despite their ineffectiveness, we included them as a special major category for reference.
\end{itemize}

\rev{The first three categories—\textbf{Version Adjustment}, \textbf{Patching Vulnerabilities}, and \textbf{Using Alternative Secure Libraries}—are typically addressed before building projects, while the remaining categories can be applied both pre-build and at runtime.
As shown in Figure~\ref{fig:taxo}, \va, \patch, and \chgconfig together account for 72.63\% of issues with solutions, reflecting a community preference for their straightforward implementation. Although these common solutions—upgrading, patching-are frequently supported by SOTA tools (as discussed in Section~\ref{sec: tool study}), other \rs, such as migration, configuration, bypassing, etc., still represent 50.75\% of actionable solutions. This significant proportion suggests that many community-driven remediation approaches are not yet supported by existing SOTA tools.
Note that constructing a taxonomy without any category overlap is inherently challenging. To address this, we employed a voting mechanism to collectively define categories and minimize overlaps. For example, we strictly define \textbf{Bypass} as a subtractive operation, while \textbf{AltLib} involves replacing a library rather than merely removing it. In practice, ambiguous cases were few, and inconsistent cases primarily occurred in minor categories instead of major ones (3.06\%), ensuring minimal impact on conclusions. 
}

\mybox{
\textbf{Finding 1}: \rev{We identified six major categories of \rs and constructed a hierarchical taxonomy, revealing that \va, \patch, and \chgconfig collectively account for the majority (72.63\%) of actionable \rs, reflecting a community preference. Notably, approximately 50.75\% of actionable solutions are commonly not supported by SOTA remediation tools. Regarding the remediation approach, most potentially feasible solutions are overlooked by existing SOTA tools, despite their potential to significantly enhance the security of vulnerable software.}
}

\subsubsection{\rev{Generalizability of \rs Distribution across Multiple Factors}}
\label{sec:generalization}
\rev{Since the adoption rates of \rs may vary due to contextual factors such as project scale, complexity, governance, and type, we conducted an analysis of the correlation between \rs adoption rates and various project characteristics. To generalize our conclusions across different factors, we first decomposed these factors into measurable parameters to verify whether the foundational conclusion—the adoption rates of \rs—remains consistent across diverse contexts.
The parameters used for this analysis are outlined in Table~\ref{tab:generalization} and include: KLOC (thousands of lines of code), number of commits, number of contributors, number of programming languages, average first response time for all issues that receive comments, average response rate for all issues, and the classification of the project as either an application or a library.
The classification of repositories into Applications (directly used by end users) or Libraries (imported to other software) was determined through a majority voting scheme conducted by the first three authors, based on an examination of the \texttt{README.md} files. In addition to applications and libraries, which account for 92\% of the repositories, we also labeled repositories primarily used for storing various types of data as \textit{data} (1\%). The remaining repositories did not contain a \texttt{README.md} file and could not be reliably categorized.}

\rev{In Table~\ref{tab:generalization}, we present two commonly used metrics to measure distribution correlation: \textbf{Pearson Correlation}~\cite{pearsoncorrelation} (shown on the left, where a value of 1 indicates the highest correlation) and \textbf{Jensen-Shannon Divergence}~\cite{hoyososorio2024representationjensenshannondivergence} (shown on the right, where a value of 0 indicates the highest correlation). 
For each parameter, issues were sorted, and a 95\% confidence interval was established by excluding both extremes of the distribution. The remaining data was then divided into four quartiles, with the deviations of each quartile from the overall dataset highlighted. Notably deviated values are bolded in the table for clarity.
The last quartile of the \textbf{KLOC} parameter exhibits the largest deviation, as the adoption rate of \texttt{patch} is significantly higher than in the other quartiles. This indicates that large-scale projects tend to favor the more universal \rs, such as \texttt{patch}, likely due to its minimal impact on the rest of a large codebase.}

\rev{The deviation observed in the first quartile of the response time parameter is attributed to the disproportionately high proportion of \( \texttt{va} \), which accounts for the rapid response to issues. This can be explained by the ease and speed with which \( \texttt{va} \) \rs can be adopted.
In terms of Application and Library deviations, issues originating from Applications show significant deviation, whereas Libraries exhibit little to no deviation—likely because most issues are concentrated within Libraries. For Applications, the distribution of all \rs deviates from the overall distribution, with exceptionally high occurrences of \( \texttt{reinit} \) and \( \texttt{chgconfig} \). This suggests that Applications, often designed with extensive user-configurable options, tend to benefit from a distinct set of \rs compared to Libraries.}

\begin{table}[]
\centering
\scriptsize
\caption{\rev{Correlation of Quartiles Across Various Factors based on Pearson Correlation and JS Divergence}}
\setlength{\tabcolsep}{5pt}
\label{tab:generalization}
\begin{tabular}{@{}l|rrrrrrr@{}}
\toprule
{\color[HTML]{000000} }                                    & \multicolumn{2}{c|}{{\color[HTML]{000000} \textbf{Scale}}}                                                  & \multicolumn{2}{c|}{{\color[HTML]{000000} \textbf{Complexity}}}                                                              & \multicolumn{2}{c|}{{\color[HTML]{000000} \textbf{Governance}}}                                                    & \multicolumn{1}{l}{{\color[HTML]{000000} }}                                               \\ \cmidrule(lr){2-7}
\multirow{-2}{*}{{\color[HTML]{000000} \textbf{Quartile}}} & {\color[HTML]{000000} \textbf{KLOC}}        & \multicolumn{1}{l|}{{\color[HTML]{000000} \textbf{\#Commit}}} & {\color[HTML]{000000} \textbf{\#Contributor}}              & \multicolumn{1}{l|}{{\color[HTML]{000000} \textbf{\#Language}}} & {\color[HTML]{000000} \textbf{Response Time}} & \multicolumn{1}{l|}{{\color[HTML]{000000} \textbf{Response Rate}}} & \multicolumn{1}{l}{\multirow{-2}{*}{{\color[HTML]{000000} \textbf{App(1st) / lib(2nd)}}}} \\ \midrule
\rowcolor[HTML]{EFEFEF} 
{\color[HTML]{000000} 1st}                                 & {\color[HTML]{000000} 0.95/0.051}           & \cellcolor[HTML]{EFEFEF}{\color[HTML]{000000} 0.945/0.055}    & \cellcolor[HTML]{EFEFEF}{\color[HTML]{000000} 0.923/0.061} & {\color[HTML]{000000} 0.923/0.062}                              & {\color[HTML]{000000} \textbf{0.877/0.079}}   & {\color[HTML]{000000} 0.951/0.046}                                 & {\color[HTML]{000000} \textbf{0.821/0.095}}                                               \\
{\color[HTML]{000000} 2nd}                                 & {\color[HTML]{000000} 0.947/0.054}          & {\color[HTML]{000000} 0.948/0.052}                            & {\color[HTML]{000000} 0.917/0.061}                         & {\color[HTML]{000000} 0.906/0.07}                               & {\color[HTML]{000000} 0.891/0.072}            & {\color[HTML]{000000} 0.947/0.05}                                  & {\color[HTML]{000000} 0.954/0.048}                                                        \\
\rowcolor[HTML]{EFEFEF} 
{\color[HTML]{000000} 3rd}                                 & {\color[HTML]{000000} 0.947/0.049}          & \cellcolor[HTML]{EFEFEF}{\color[HTML]{000000} 0.95/0.048}     & \cellcolor[HTML]{EFEFEF}{\color[HTML]{000000} 0.944/0.054} & {\color[HTML]{000000} 0.962/0.041}                              & {\color[HTML]{000000} 0.95/0.052}             & {\color[HTML]{000000} 0.931/0.06}                                  & {\color[HTML]{000000} N.A.}                                                               \\
{\color[HTML]{000000} 4th}                                 & {\color[HTML]{000000} \textbf{0.867/0.079}} & {\color[HTML]{000000} 0.902/0.075}                            & {\color[HTML]{000000} 0.925/0.064}                         & {\color[HTML]{000000} 0.971/0.038}                              & {\color[HTML]{000000} 0.961/0.042}            & {\color[HTML]{000000} 0.929/0.058}                                 & {\color[HTML]{000000} N.A.}                                                               \\ \bottomrule
\end{tabular}

\end{table}

\subsection{RQ2: Acceptance of \RS}
\label{sec:rq2}
\ly{Given the \rs categories, we further explored their acceptance regarding individual categories by manually going through these issues. Next, the reasons behind the rejected \rs were also investigated and summarized.
}

\subsubsection{Analysis of Acceptance of \RS}
The Close Card Sorting~\cite{hybridcardsorting} with split three folds was leveraged to label the acceptance due to the pre-defined categories. 
\rev{Since GitHub issues can be arbitrarily closed by maintainers without explicit reasoning, an \texttt{unknown} label was incorporated into our set of candidate labels. The final few comments in an issue discussion often serve as closing discussions and conclusion remarks. These comments typically provide critical hints regarding the acceptance of RT. Specifically, in addition to analyzing the title and initial comment, we reviewed comments in reverse order from the end to identify those that determine the acceptance status until a comprehensive understanding of the discussion trend is comprehended. Overall, over 90\% of accept/reject issues could be confirmed within the last 5 comments.} 
In the case of positive and negative feedback, it was labeled as \texttt{accept} and \texttt{reject} respectively. If there was no relevant feedback, \texttt{unknown} was assigned.

\ly{In total, $15,837$ issues had their \rs accepted and $4,980$ were rejected, ignoring the \texttt{unknown} label. Because only closed GitHub issues were involved, the issues with accepted \rs were more likely to be included than those with rejected \rs. The major adopted \rs were \patch (23.41\%), \chgconfig (19.72\%), and \va (19.59\%), which generally follow a similar distribution in all issues. }

\subsubsection{Audience Identity Determination}
\label{sec:party}
\ly{During the labeling, we found that the audiences of \rs in each issue were not consistent and might exhibit different identities, such as the maintainers and the downstream users of the repositories. To avoid biasing the acceptance with blended audiences, we first identified the identities of the audiences and then separately summarized the acceptance for finer granularity. As we found downstream users and maintainers are predominant identities, we labeled the audiences of \rs as \texttt{User}, \texttt{Maintainer}, \texttt{Other}, and mainly analyzed the former two.}

\ly{
During the previous Close Card Sorting~\cite{hybridcardsorting}, we also labeled the identities of \rs audiences from the three categories. (1) First we collected all contributors registered in each GitHub repository and saved them for reference. (2) For the comments before issue closing, we identified the relevant feedback and then collected the usernames of those commenters. (3) We cross-checked if the commenters who provided the feedback for \rs are from the corresponding contributors in the current repository. If yes, this issue would be labeled as \texttt{Maintainer}. (4) If not, the annotator further checked if the commenters were using the repositories from the context, especially from the initial comments. Otherwise, it would be considered \texttt{User}.
\rev{There were cases we considered out of scope because they do not align with the existing classifications, thus labeled as \texttt{Other}: (1) Invalid Comments: These are comments that do not directly respond to the initial creation comments or are made by individuals who are neither maintainers nor users, such as other users merely following the issue.
(2) Invalid Commentors: The commentators are bots or individuals whose identities cannot be determined from their comments.
}
Ultimately, out of the total 21,187 issues, 7,427 (35.05\%) ones were classified as \texttt{Maintainer} and 13,763 (64.96\%) ones as \texttt{User}.
}

\rev{We take an example~\cite{githubexample1} in Section~\ref{sec:categorization} to demonstrate how the identity of the audience is determined. From the intention described in the creation comment in the quoted text above and the reply from the repository maintainer in the comment below, we can deduce that the issue creator provided a solution to the vulnerability, which was subsequently accepted by the maintainer. Thus, the audience identity should be \texttt{Maintainer}.
}

\begin{figure}[ht!]
    \centering
  \includegraphics[width=0.80\linewidth]{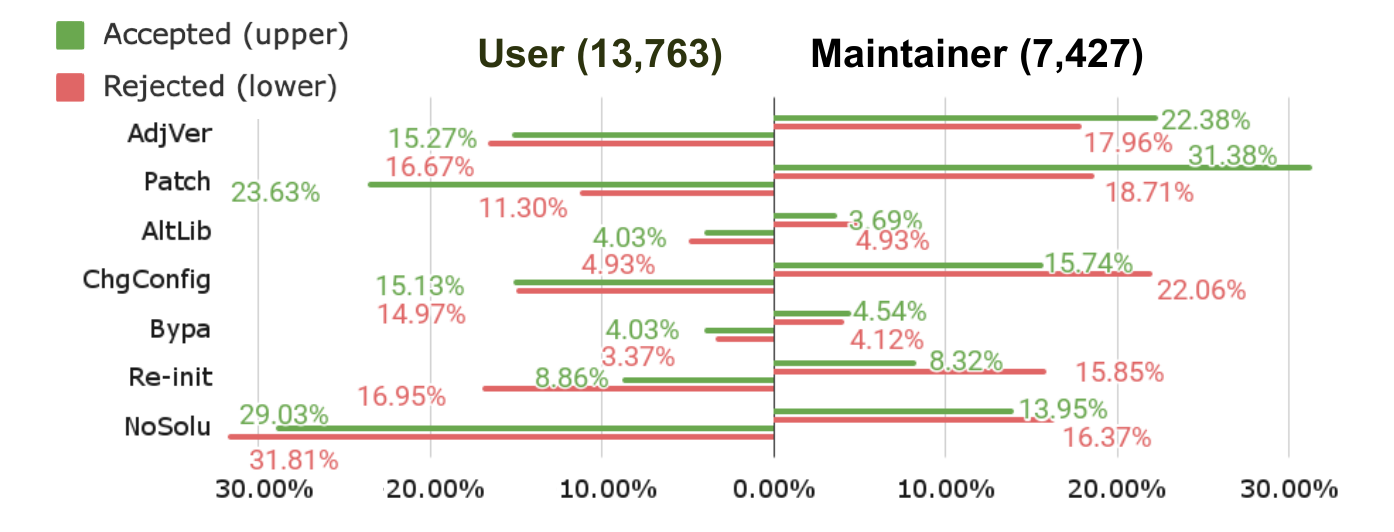}
  \caption{Distribution of \rs towards Acceptance and Audience of Issues. \textbf{Proportions are normalized within each combination. For example, the sum of all  Accept-User green bars on the left is 100\%.}}
  \label{fig:accept_identity}
\end{figure}

In Figure~\ref{fig:accept_identity}, 
the rest of the dataset can be split by the acceptance of \rs (\textit{Accepted} \& \textit{Rejected}) and the identities (\textit{User} \& \textit{Maintainer}) into four combinations, which are \textit{Accepted-Maintainer} (5,877), \textit{Accepted-User} (9,960), \textit{Rejectd-Maintainer} (1,388), \textit{Rejected-User} (3,593).  The distribution of \rs is depicted in four scenarios. As the numbers in each combination vary greatly, they are normalized to proportions within each combination for a fair comparison. The left side of the mirror chart is the issues of  \sk, and \pv is on the right. 
\rev{Overall, for \sk, \patch is the most favored tactic, as indicated by a significantly higher acceptance rate compared to its rejection rate, while other categories show no notable differences in acceptance rates. In contrast, for \pv, \va and \patch are the most commonly adopted tactics, together accounting for over 50\% of maintainer-accepted solutions. This highlights a clear preference for these \rs, emphasizing the need for greater support from modern tools.}


\mybox{
\textbf{Finding 2}: \rev{The notable discrepancy revealed by the finding that 64.96\% of participants are \texttt{user}, while only 35.04\% are \texttt{maintainer} in security issue discussions, highlights a significant demand for security expertise. This imbalance emphasizes a crucial gap that SOTA tools have yet to effectively address, underscoring the need for more comprehensive solutions that cater to a wider range of security requirements.
}}

\begin{figure}[h]
    \centering
    \begin{minipage}{0.5\textwidth}
        \centering
        \includegraphics[width=\textwidth]{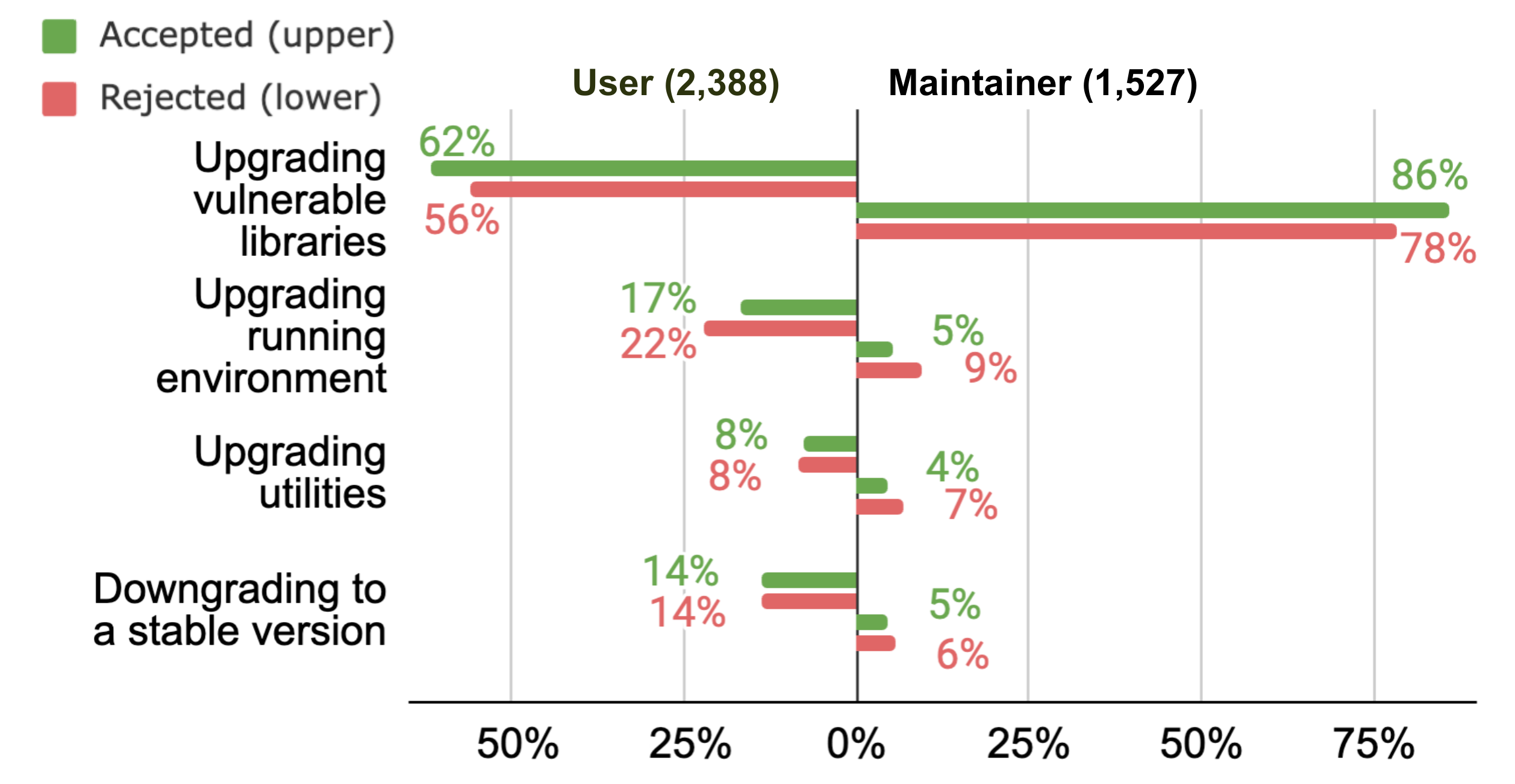}
        \caption{Subcategories of Adjusting Versions}
        \label{fig:upgrade}
    \end{minipage}\hfill
    \begin{minipage}{0.5\textwidth}
        \centering
        \includegraphics[width=0.95\textwidth]{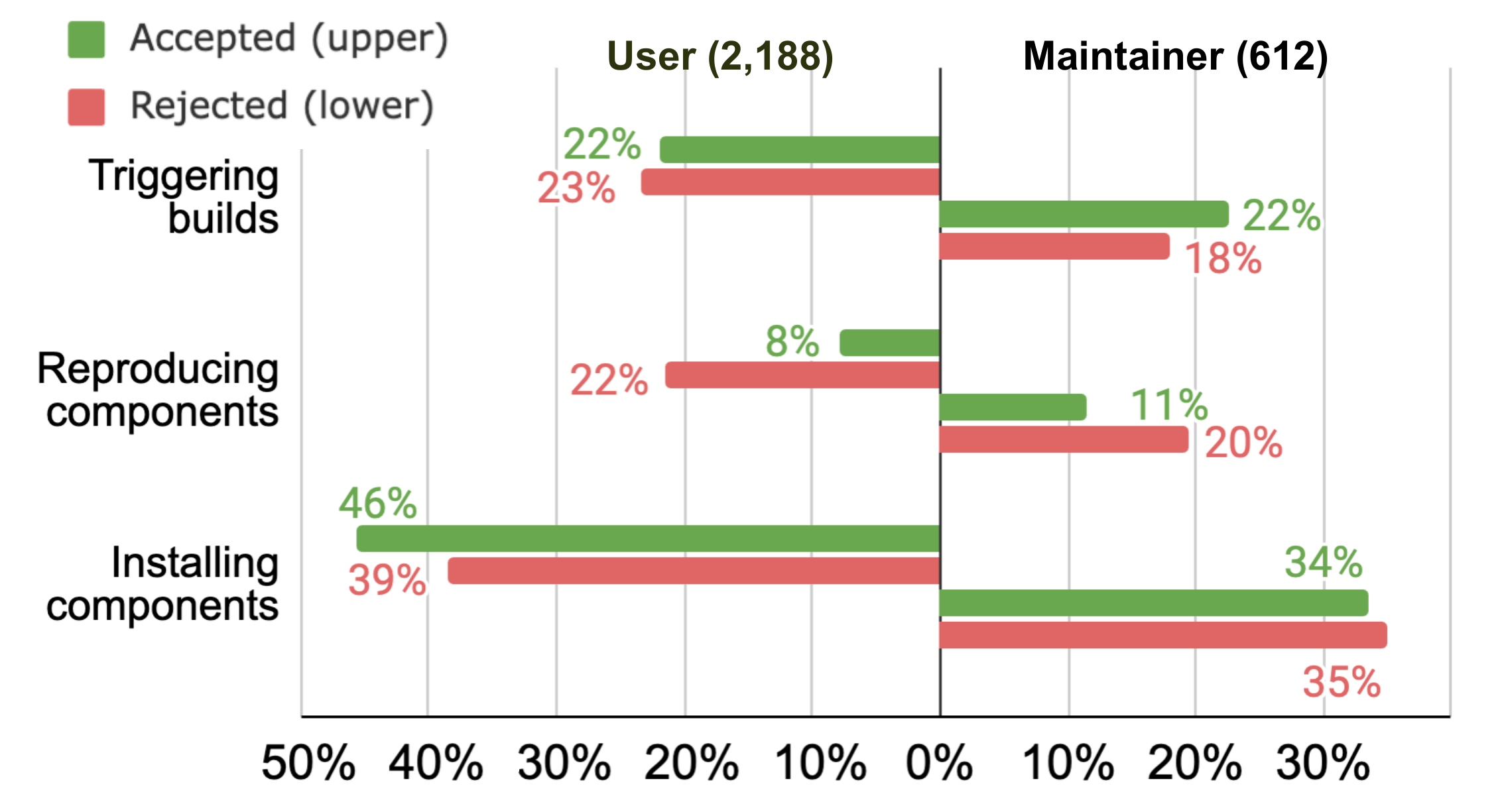}
        \caption{Subcategories of Re-initializing}
        \label{fig:initialize}
    \end{minipage}
\end{figure}

Individual category analysis is elaborated as follows:  
\begin{itemize}[leftmargin=5pt]    
    \item For \va, \rs have a larger accepted ratio of \rs for \pv than the rejected. It is plausible that these categories are more readily accepted by repository maintainers because the changes they entail, such as version adjustments in BOM, are typically minor and follow uniform formats, making them easier to review. While the accepted and rejected \rs for \sk are not evidently different, after exploring the reason, two main reasons were found: \rev{\ding{172} \sk (mostly project users) often lack solid testing and open reviews to prevent potential breaking changes caused by version adjustments.
    \ding{173} \sk are not obligated to resolve issues promptly, with many opting to wait for upstream fixes, thus being categorized under \nosolu. The acceptance rate for \nosolu is notably higher at 29\% compared to 15\% for \va.
    We further break down the distribution of \va subcategories in Figure~\ref{fig:upgrade}. \textit{Upgrading vulnerable libraries} holds the largest share and is generally adopted more frequently than other tactics. In contrast, \textit{Upgrading the running environment}—which involves updating compilers, containers, or operating systems—can introduce unexpected changes and requires significant adaptation efforts, making it less favored by both \sk and \pv.}
    
    \mybox{
    \textbf{Finding 3}: 
    \rev{Surprisingly, \pv exhibits a preference for \va, whereas \sk does not demonstrate a similar tendency. This is because \sk, primarily composed of project users, lacks reliable mechanisms to mitigate potential breaking changes introduced by adjusted versions. Consequently, a significant portion of \sk participants prefer to wait for upstream updates rather than upgrading to sub-optimal versions.
    The preferences of different audiences indicate the potential need for customized remediation suggestions from future remediation tools. 
    }
    }
    
    \item \patch is seen to have evidently higher accepted ratios for both \sk and \pv. Furthermore, \patch has the outstanding accepted ratio over all other major categories, indicating the most welcomed strategy. This could be attributed to the predictable impact of diff in code snippets, which is much easier to review than numerous changes by other \rs. Moreover, we counted the number of changed lines in commits mentioned in the events of issues, the median is 216.3, indicating relatively small-sized changes involved.
    \item The proportions of both accepted and rejected \rs do not have distinguishing differences towards \altlib, and \bypass. Hence, regardless of the identities of the audience of \rs, the acceptance of \rs is not likely to be different for them. The low proportions of certain categories, like \altlib, are influenced by the availability of suitable alternative libraries. Consequently, acceptance rates vary significantly, as different alternative libraries require varying levels of adaptation effort, unlike more predictable efforts of \va.

    \item \chgconfig is not preferred by \pv as indicated by the 7\% difference of acceptance ratios, while the difference is trivial for \sk. 
    We further investigated the proportions of operations under \chgconfig and found that \textit{Disabling features} was rarely accepted because either it may disrupt other functionalities supported by the features or the feature is potentially used in certain circumstances. Therefore, \rs subjects to contextual scenarios are more likely to be rejected. 

   
    \item \textbf{\reinit} is not favored by both \pv and \sk, with the acceptance ratio of \rs nearly halved by rejections. It is reasonable because \reinit is usually temporary without lasting effects, typically useful for short-term measures. For further analysis, in Figure~\ref{fig:initialize}, several main operations of \reinit are displayed. 
    It shows that \sk prefers \textit{Installing components} and tends to avoid \textit{Reproducing and verifying components}, indicating a preference for simpler procedures over those requiring intensive efforts. Contrastingly, \pv tends to be conservative about installing new components, as this can lead to increased maintenance efforts in the long run. 
    It is observed that \sk, as users, often seek quick and effective measures without considering the future maintenance burdens they may introduce. In contrast, maintainers, who are responsible for the long-term upkeep of OSS repositories, are reluctant to implement major changes without thorough validation.
\end{itemize}

    \mybox{
\rev{\textbf{Finding 4}: Due to the relatively small size of patches (216.3 median LoC), \patch is generally well-received. In contrast, tactics such as \altlib, \bypass, and \chgconfig, where effectiveness depends on specific factors like the availability of alternative libraries, tend to have lower acceptance rates. 
Insights from \reinit reveal that \sk favors lightweight tactics, and \pv remains cautious about introducing new components, as they may lead to long-term maintenance challenges. Given the complexities of \rs and client preferences, remediation tools can enhance effectiveness by tailoring suggestions to specific contexts.}
    }

\subsubsection{Reasons Behind Rejected \RS}
\label{sec:reasonsrejection}
Next, the reasons why \rs were rejected are explored to promote future research on resolving the concerns of audiences. As the rejection reasons of \rs may depend on the identities of the audience, i.e. issue creators and repository maintainers, in this subsection, the reasons were also summarized into categories based on \pv and \sk respectively. Due to the diversity of categories of rejection reasons, we conducted another Hybrid Card Sorting with Cross-review following the same manner as in Section~\ref{sec:categorization}.
$3,241$ issues were excluded because either the rejection was not justified ($3,014$) or the issues were redirected (227). Eventually, 17 categories of reasons were summarized from 1,486 issues.


\rev{We take an example to illustrate the process of rejection reason summarization. The case~\cite{githubexample2} was collected from the dataset. The quoted text below demonstrates the reply from a maintainer who reviewed the proposed \rs. Based on all subsequent comments, we infer that the proposal was ultimately rejected. 
The rejection reason is attributed to the updating of PKG\_RELEASE, which caused unexpected changes and led to a failed Docker build. Consequently, the reason is categorized as: "The fix disrupts existing functionality."
}

\begin{shadedquotation}
\footnotesize
\rev{Quoting from an example Issue~\cite{githubexample2}: \\
I took a look at opening a PR to change the stable images to 1.24.0-r7, but updating PKG\_RELEASE to 7 caused some unexpected changes that made the rest of the docker build fail. Configuration directories/files seem to have been relocated between these releases, so it seems like it might be a breaking change for many users.}
\end{shadedquotation}

\rev{As shown in Table~\ref{tab:reject}, the 17 categories are divided into two groups: \textbf{Breaking} and \textbf{Non-Breaking}. \textbf{Breaking} tactics refer to those that disrupt system functionality, rendering it unusable, while \textbf{Non-Breaking} tactics maintain usability but may introduce other concerns.
Breaking issues significantly outnumber non-breaking ones for both \sk and \pv roles. However, \pv demonstrates greater concern for non-breaking issues, as these can gradually increase the maintainability burden or introduce other long-term challenges. Given the open nature of OSS repositories, maintainers are naturally attentive to both breaking and non-breaking issues.
From these reasons, it is evident that a usable \rs should fully resolve security issues without disrupting existing functionalities or introducing new vulnerabilities. To improve acceptance, it is beneficial to provide a threat evaluation and compare the \rs with alternative solutions to emphasize its necessity and advantages. Simplifying the \rs procedures and offering supporting evidence can further enhance its effectiveness.
}


\mybox{
\rev{\textbf{Finding 5}: 
Overall, breaking issues significantly outnumber non-breaking ones. Compared to \sk, \pv places 10\% more emphasis on non-breaking issues due to their impact on maintainability.
For \rs to be accepted, they must fully resolve issues while preserving functionality, maintainability, and security. Audiences also consider the required effort, threat evaluations, and comparisons with alternatives. These diverse concerns underscore the complexity of client needs and suggest potential areas for optimizing future remediation tools.}
}

\begin{table}[]
\caption{Reasons of Rejected \RS}
\scriptsize
\label{tab:reject}
\setlength{\tabcolsep}{3pt}
\begin{tabular}{@{}lrrrlrrr@{}}
\toprule
\textbf{Reason (Breaking)} & \multicolumn{1}{l}{\textbf{Count}} & \multicolumn{1}{l}{\textbf{User}} & \multicolumn{1}{l}{\textbf{Maintainer}} & \textbf{Reason (Non-breaking)} & \multicolumn{1}{l}{\textbf{Count}} & \multicolumn{1}{l}{\textbf{User}} & \multicolumn{1}{l}{\textbf{Maintainer}} \\ \midrule
Fix fails to completely address issues  & 329                                & 25\%                              & \multicolumn{1}{r|}{16\%}               & Fix is not the best solution                & 160                                & 10\%                              & 12\%                                    \\
Fix is out of scope                     & 309                                & 21\%                              & \multicolumn{1}{r|}{19\%}               & Fix duplicates existing functionality       & 81                                 & 4\%                               & 9\%                                     \\
Fix disrupts existing functionality     & 148                                & 7\%                               & \multicolumn{1}{r|}{17\%}               & Fix requires significant effort             & 81                                 & 5\%                               & 6\%                                     \\
Fix is unrelated to the issue           & 110                                & 9\%                               & \multicolumn{1}{r|}{5\%}                & Fix introduces a new issue                  & 68                                 & 4\%                               & 5\%                                     \\
Fix is risky                            & 55                                 & 4\%                               & \multicolumn{1}{r|}{3\%}                & Fix is based on insufficient evidence       & 38                                 & 3\%                               & 2\%                                     \\
Fix requires unavailable resources      & 40                                 & 3\%                               & \multicolumn{1}{r|}{2\%}                & Fix compromises maintainability             & 22                                 & 1\%                               & 2\%                                     \\
Fix relies on insecure dependencies     & 31                                 & 2\%                               & \multicolumn{1}{r|}{1\%}                & Fix is too late                             & 7                                  & 0\%                               & 1\%                                     \\
Fix is identified as a mistake          & 2                                  & 0\%                               & \multicolumn{1}{r|}{0\%}                & Fix is unethical                            & 3                                  & 0\%                               & 0\%                                     \\
Total in this category                  & 1,024                              & 73\%                              & \multicolumn{1}{r|}{63\%}               & Fix is flaky                                & 2                                  & 0\%                               & 0\%                                     \\
-                                       & -                                  & -                                 & -                                       & Total in this category                      & 462                                & 27\%                              & 37\%                                    \\ \bottomrule
\end{tabular}
\end{table}

\subsubsection{\rev{Comparison Study on Non-highly Starred Repositories}}
\label{sec:lowstar}
\begin{figure}[]
    \centering
    \begin{minipage}{0.45\textwidth}
        \centering
        \includegraphics[width=\textwidth]{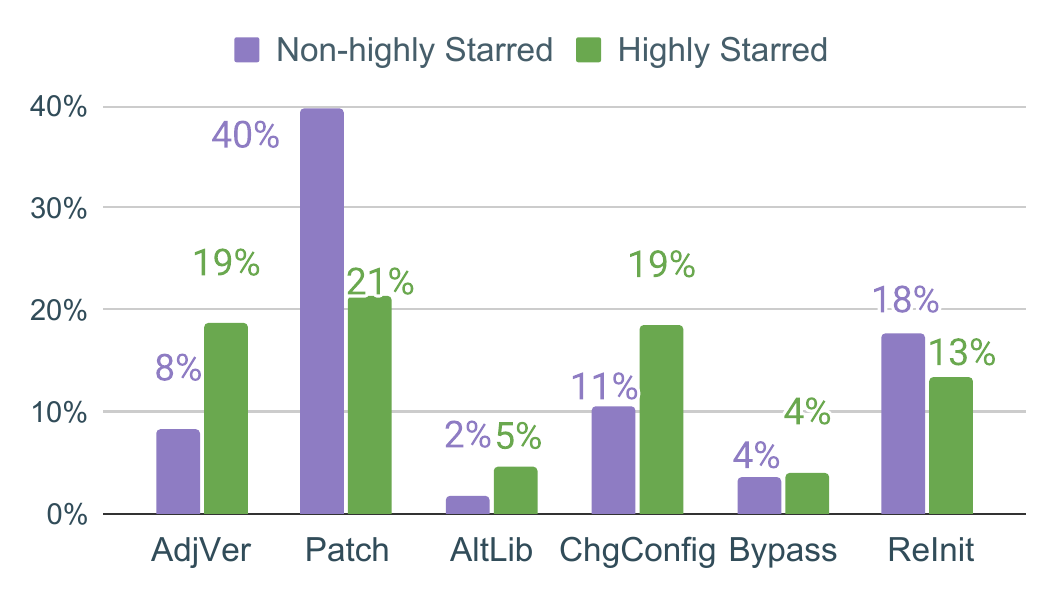}
        \caption{\rev{Distributions of \rs against Repo Stars}}
        \label{fig:lowstar}
    \end{minipage}\hfill
    \begin{minipage}{0.45\textwidth}
        \centering
        \includegraphics[width=\textwidth]{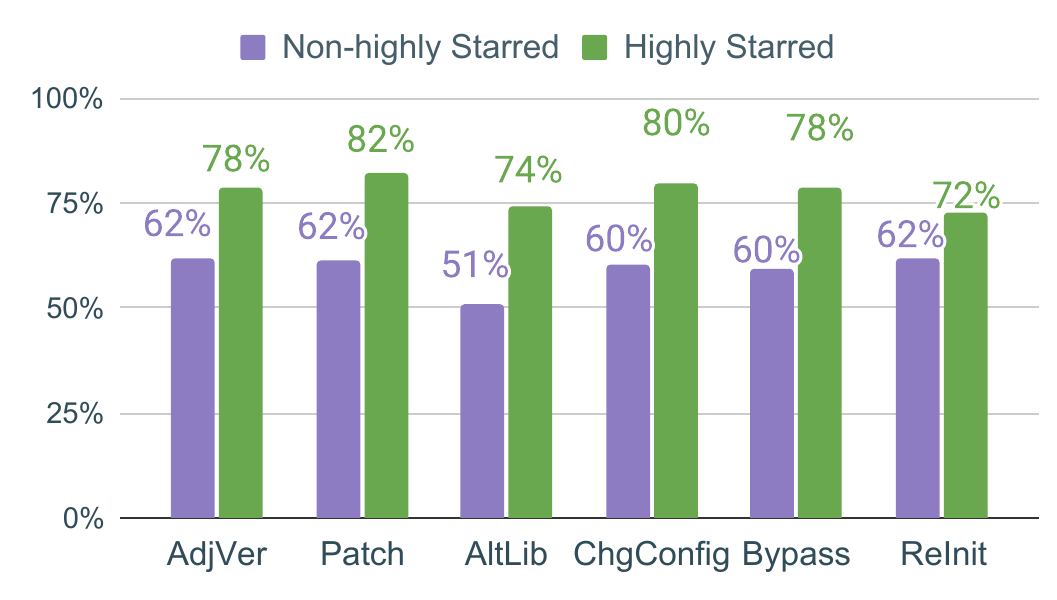}
        \caption{\rev{Acceptance of \rs against Repo Stars}}
        \label{fig:lowstaraccept}
    \end{minipage}
\end{figure}

\rev{Since our focus is on repositories with high-quality \rs, our previous findings were derived from issues in highly starred repositories. To assess the generalizability of our conclusions to repositories with fewer stars, we conducted a supplementary study on repositories with fewer than 10k stars. We randomly selected 3k issues to analyze their \rs, distributions, and acceptance rates.
In Figure~\ref{fig:lowstar}, the distribution of \rs across major categories is illustrated. The High Star dataset refers to the one collected in Section~\ref{sec:data}.
}

\rev{We found that \patch is significantly more prevalent in repositories with lower star counts, possibly due to the relatively smaller code corpus. Consequently, other \rs exhibit lower adoption rates, particularly \va and \chgconfig, which may reflect less stable version and dependency management, as well as less mature configurations in non-highly starred repositories.
Notably, we identified an uncollected \rs operation, overriding transitive dependencies~\cite{mavendepmanage}, which falls under the \va category. This operation is typically considered a temporary solution and is not recommended for long-term projects~\cite{overridetransitivedep1, overridetransitivedep2,zhang2023mitigating}. As a result, it is uncommon in highly starred repositories, which usually have more mature issue resolution mechanisms.
}

\rev{
In Figure~\ref{fig:lowstaraccept}, the acceptance rates for all \rs are generally higher in highly starred repositories. This may be attributed to the relatively higher quality of \rs and more in-depth discussions in these repositories. Despite the high adoption rate of \patch in non-highly starred repositories, its acceptance rate remains similar to other \rs, with overall acceptance rates being relatively close across all \rs. Overall, while the distribution and acceptance rates of \rs in non-highly starred repositories differ, the relative trend remains consistent.}

\subsection{RQ3: Cost Analysis of Security Issues}
To further assess the cost of \rs, we analyzed the time and human resources required for each issue, providing insights into both cost and effectiveness. This analysis offers a comprehensive view of the issue's lifecycle and human interactions. We use time duration as an indirect metric for \rs cost, a widely adopted approach in recent empirical studies~\cite{hu2024empirical,pan2024unveil}.


\begin{figure}[]
    \centering
    \begin{minipage}{0.4\textwidth}
        \centering
        \vspace{0.1cm}
        \includegraphics[width=0.6\textwidth]{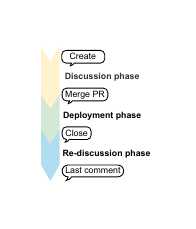}
        \vspace{0.5cm}
        \caption{Issue Timeline and \\Critical Checkpoint}
        \label{fig:timeline}
    \end{minipage}\hfill
    \begin{minipage}{0.6\textwidth}
        \centering
        \includegraphics[width=0.9\textwidth]{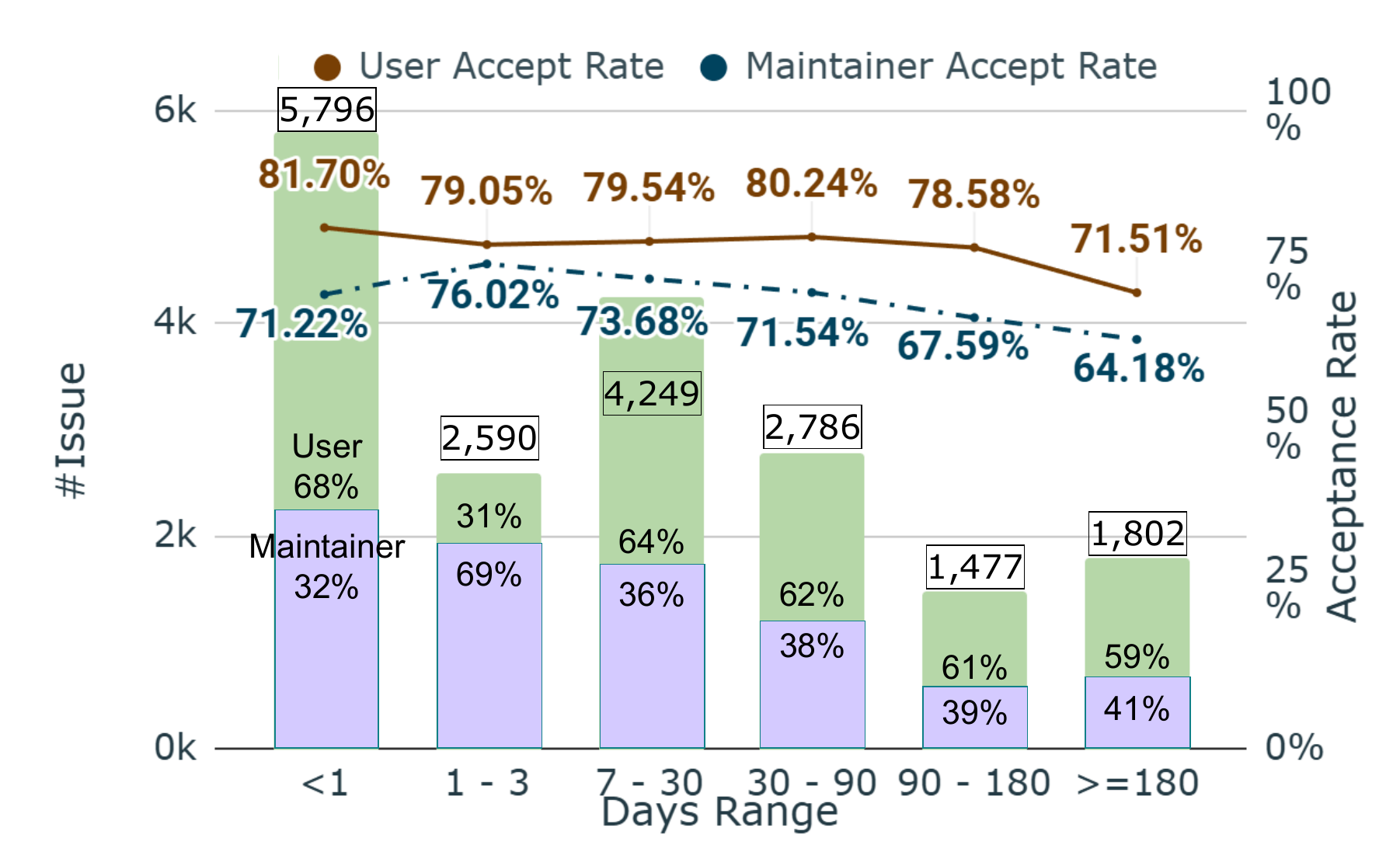}
        \caption{Duration and Acceptance Rates \\for Discussion Phase}
        \label{fig:timeline_chart}
    \end{minipage}
\end{figure}

The timeline of an issue with critical checkpoints is illustrated in Figure~\ref{fig:timeline}. As we only analyzed closed issues, the checkpoints always involve \textit{close}. The phase from \textit{Creation} to \textit{Merge PR} is denoted as \textit{Discussion} phase. Optionally, if the remediation involves file changes, there could be a PR merging mentioned as an event along the timeline before \textit{close}. After merging PR, the phase is denoted as \textit{Deployment}. Notably, there exist quantities of issues that either are reopened or have comments created after \textit{close} even for a long time according to our observation. Thus, we included \textit{last comment} as one of the checkpoints which mark \textit{Re-discussion} phase after closing the issue. The \textit{Re-discussion} are labeled if any condition of two is met: (1) The issue is reopened. (2) The last comment is created after the issue is closed. As for the implementation, from the crawled timelines, we iteratively went through the events and extracted the checkpoints based on the event names.

\subsubsection{Discussion Phase}
The duration of the \textit{Discussion phase} was quantitatively measured and demonstrated in Figure~\ref{fig:timeline_chart}. The median for \textit{Discussion phase} is 6.1 days, and the average is 50.76 days. It is observed that $5,796$ ($30.99\%$) issues are closed in one day, and 6,065 ($32.43\%$) of issues remained open for over 30 days, providing a prolonged window for attackers, which aligns with findings from a study~\cite{pan2024unveil} that $32.1\%$ of vulnerabilities have attack windows exceeding 30 days. 
\rev{The acceptance rate of \sk consistently surpasses that of \pv, suggesting that repository maintainers apply higher standards when accepting remediation suggestions. Additionally, overall acceptance rates tend to decline as the \textit{Discussion phase} lengthens, reflecting the increased complexity of issues that require extended resolution time. Over time, the proportion of issues created by \sk gradually falls below 60\%, indicating that issues handled by maintainers typically demand longer review periods.
To accelerate issue resolution with viable and acceptable \rs, future automated remediation tools incorporating diverse tactics could help shorten the attack window, thereby enhancing security.}


\mybox{
\textbf{Finding 6}: \rev{A total of $28.64\%$ of issues remained open for more than 30 days, leaving an extended window of opportunity for potential attackers. The consistently higher acceptance rate of \sk compared to \pv indicates that maintainers enforce stricter criteria for approving \rs. 
Moreover, the length of the discussion phase shows a negative correlation with the acceptance rate, indicating that prolonged discussions may decrease the likelihood of \rs acceptance. To expedite issue resolution with viable and acceptable \rs, future automated remediation tools equipped with diverse tactics could help shorten the attack window, thereby improving security.}
}


\subsubsection{Re-discussion Phase}
\begin{figure}[]
    \centering
    \begin{minipage}{0.45\textwidth}
        \centering
        \includegraphics[width=\textwidth]{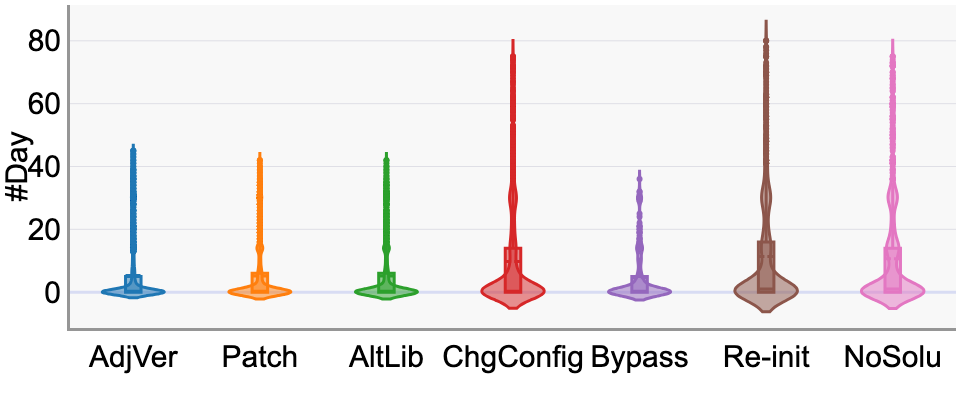}
        \caption{\rev{Violin Plot of Re-discussion Duration}}
        \label{fig:reopen_rs}
    \end{minipage}\hfill
    \begin{minipage}{0.45\textwidth}
        \centering
        \includegraphics[width=\textwidth]{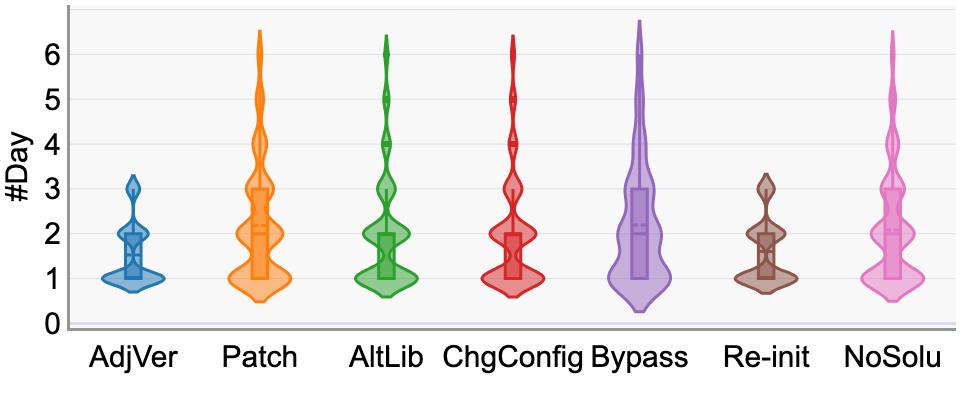}
        \caption{\rev{Violin Plot of Numbers of Contributors}}
        \label{fig:contributor_rs}
    \end{minipage}
\end{figure}

In Figure~\ref{fig:reopen_rs}, the distribution of duration of \textit{Re-discussion} over 9,648 re-opened issues is presented over major categories. The issues reopened by bots or only commented on by bots were filtered out. It is evident that \chgconfig and \reinit have much longer average duration than the rest. The extended \textit{Re-discussion} phase for these two \rs highlights the frequent reopening and prolonged discussions of issues, suggesting that they often fail to provide long-term effective solutions or are highly susceptible to evolving changes. This indicates that such measures require ongoing monitoring to effectively address new challenges over the long term if they are to be recommended as viable solutions.


\subsubsection{Reopened Issue Analysis}
\begin{figure}[]
    \centering
    \begin{minipage}{0.4\textwidth}
        \centering
        \includegraphics[width=\textwidth]{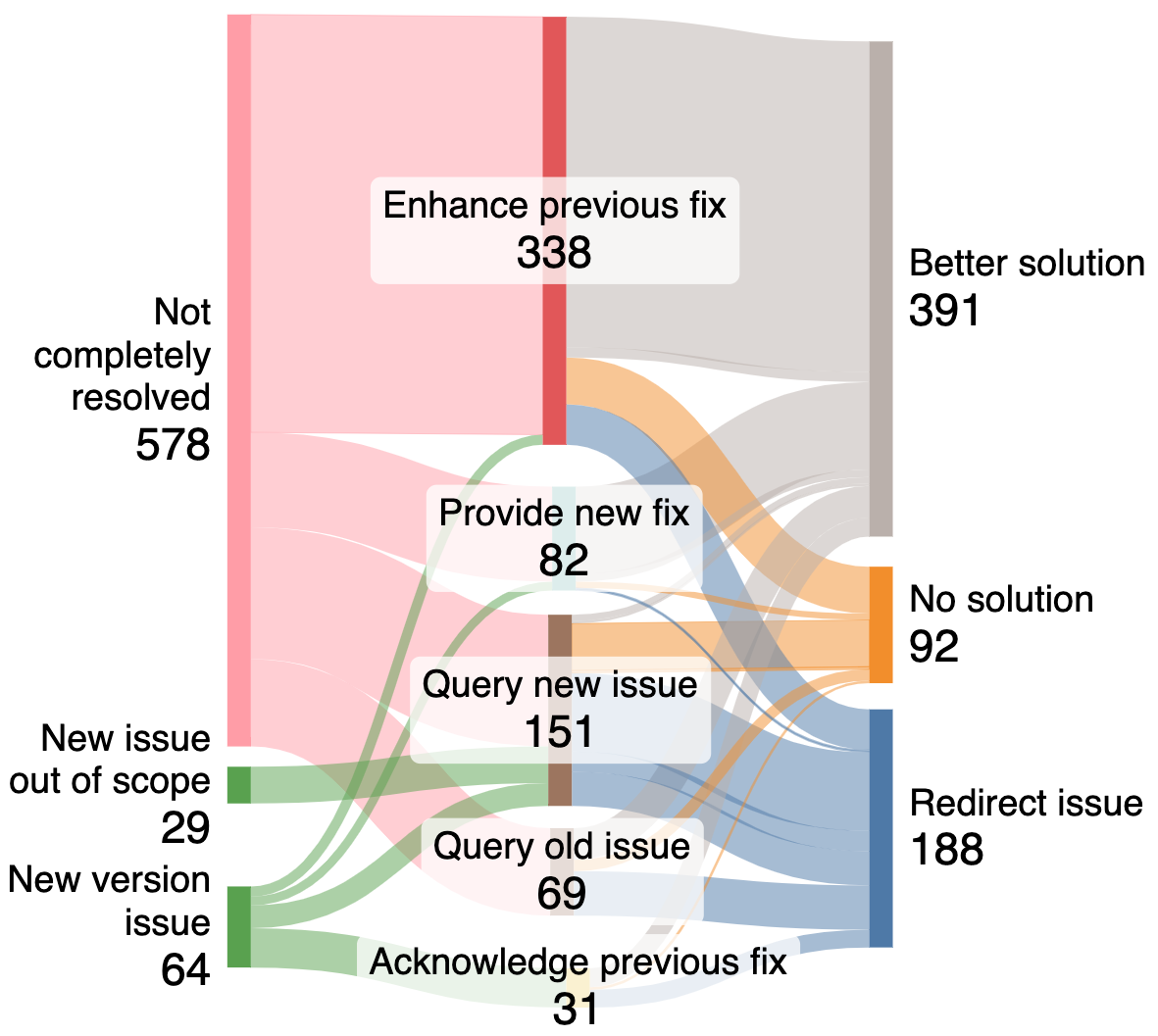}
        \caption{The Distribution of Reason, Operations, Outcome of Reopened Issues}
        \label{fig:reopen_reason}
    \end{minipage}\hfill
    \begin{minipage}{0.5\textwidth}
        \vspace{1cm}
        \includegraphics[width=\textwidth]{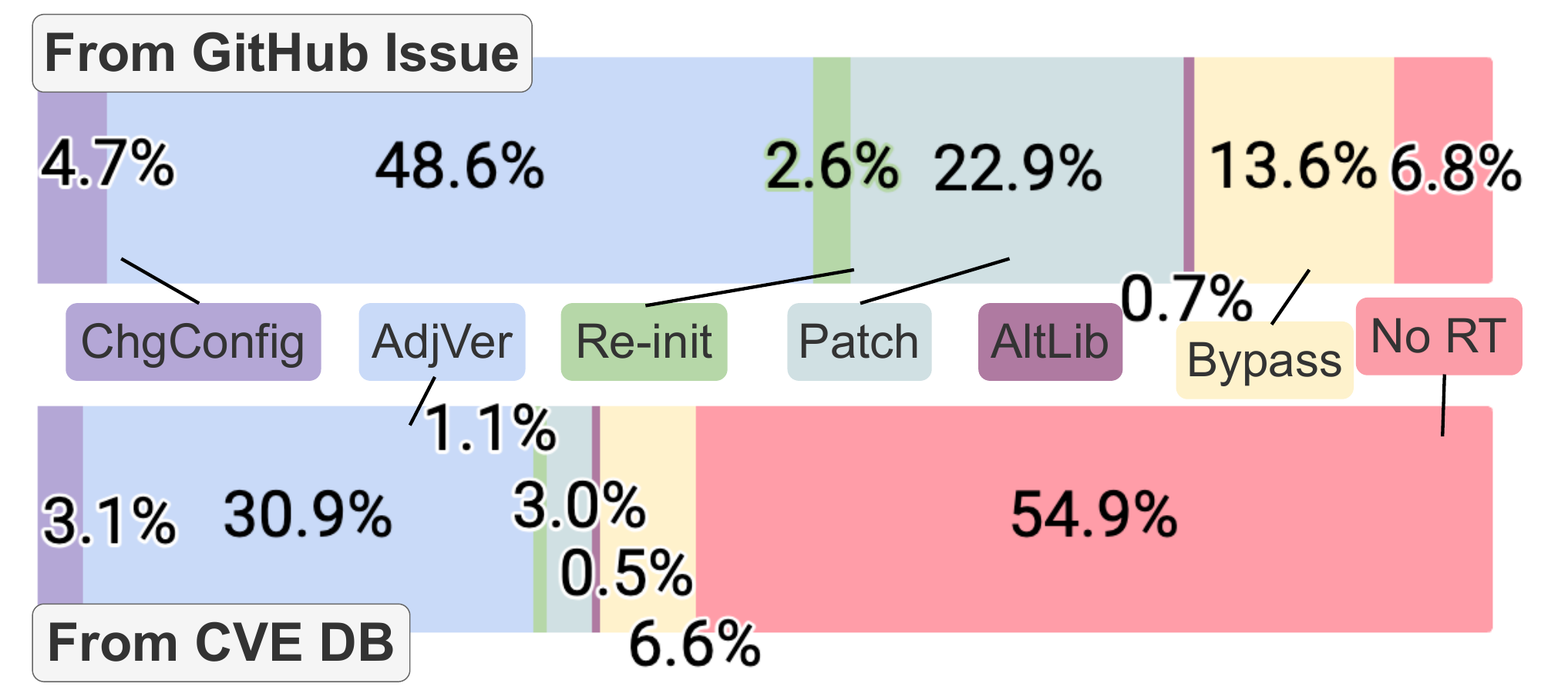}
        \vspace{0.4cm}
        \caption{Comparison of the Distributions of \rs of Vulnerability DBs and GitHub issue}
        \label{fig:advisory}
    \end{minipage}
\end{figure}

To further explore the reasons, operations, and outcomes of reopened issues, we manually went through part of them for Card Sorting. 
To ensure the issues were reopened by another thread instead of the follow-up comments of the previous threads, we conservatively selected issues with over 180 days \textit{Re-discussion phase} and issues explicitly reopened by humans. After filtering out bot-reopened issues, we obtained $778$ issues. 
If the new comments after close are irrelevant to the original issues, the issue was labeled as \textit{irrelevant} and excluded, resulting in $671$ issues in total. 
The Card Sorting conducted in the same manner as before has labeled each issue by intention, reason, and outcome as illustrated in Figure~\ref{fig:reopen_reason}. 
Notably, 578 ($93\%$) issues were reopened because they were not completely resolved previously. Peeking into the \rs categories of these 578 issues, \chgconfig and \reinit stand for $77.82\%$ out of major \rs categories, substantiating their volatility and unreliability found in the previous subsection.
Most of the issues were reopened by enhancing the previous fixes. 
Fortunately, 58\% of issues lead to better solutions as the outcomes. \textit{Redirect issue} means the commenters are redirected to other issues, which could lead to better solutions as well.

\rev{We investigate which \rs demand more human resources, measured by the number of maintainers involved in resolving security issues. Figure~\ref{fig:contributor_rs} illustrates the number of contributors engaged in each issue. This count is determined by the number of registered contributors from the corresponding repository who commented on the issue. \va and \reinit typically involve fewer contributors, as these tactics are generally easier to implement.}

\mybox{
\rev{\textbf{Finding 7}: A total of 93\% of issues were reopened due to incomplete previous fixes, with \chgconfig and \reinit identified as the most volatile \rs. Notably, 58\% of reopened issues led to improved solutions, suggesting that stakeholders in security issues could benefit from continuous monitoring to address problems arising from incomplete fixes and evolving software versions. Therefore, effective implementation of \rs should prioritize stable solutions while incorporating continuous remediation practices to enhance long-term security and reliability.
}}


\subsection{RQ4: Analysis of Remediation from Vulnerability Databases}
\label{sec:rq4}

Modern vulnerability databases (DBs) may offer various \rs, such as NVD~\cite{nvd}, Snyk Vulnerability DB~\cite{snykvuldb}, Mend (Whitesource)~\cite{mend} OSV~\cite{OSV}, GitHub Advisory~\cite{githubadvisory}, VulDB~\cite{vuldb}, VulnDB~\cite{vulndb}, Seclists~\cite{seclists}, ExploitDB~\cite{edb}, and IBM XForce~\cite{ibmxforce}. We first qualitatively explored what \rs they provide and to which extent. Then, for those DBs available to the public and free of charge, we investigated the distribution of \rs and compared them with those from previous Section~\ref{sec:rq1} to reveal the different inclinations of software security practitioners.

\subsubsection{\RS from Vulnerability DB}
We focused on universal vulnerability DBs that are not specific to certain platforms, excluding those like UbuntuDB~\cite{ubuntucve}, which focuses on Ubuntu-related issues. Additionally, DBs limited to access restrictions, such as VulnDB, were excluded. \ly{To quantitatively reveal the \rs availability in these DBs, considering that most of them have no public API for automated crawling, we could only manually scrutinize samples of entries from their websites. For the diversity, we reviewed 100 entries (5 random entries for each year in the last two decades) for each DBs leading to 900 in total. }

Table~\ref{tab:vuldb} lists the support of \rs in popular vulnerability DBs. \halfcirc\xspace means that not all entries include the feature. \emptycirc\xspace refers to no support. It is shown that only GitHub Advisory and commercial DBs have designated sections displaying remediation suggestions. Accordingly, all DBs except for ExploitDB may mention the remediation suggestions including workarounds in the vulnerability descriptions. However, the presence of \rs is not guaranteed. In fact, most of the entries have no \rs as revealed in the next subsection~\ref{sec:advisory_rs}. 
\rev{Notably, GitHub Advisory and SnykVulDB feature designated workaround sections that offer alternative solutions for users when remediation suggestions are not feasible, such that the fixed version introduces breaking changes. Additionally, only NVD, GitHub Advisory, and SnykVulDB provide dedicated links to fixes, including suggestions and patches, enabling users to quickly access external remediation resources. Overall, support for \rs remains limited in modern vulnerability databases, hindering prompt fixes for downstream users. Disclosing vulnerabilities without providing \rs not only creates a potential attack window but also fails to offer defense solutions for users, leaving a substantial gap.}
\mybox{
\rev{\textbf{Finding 8}: \Rs are inadequately supported by modern vulnerability databases. Only GitHub Advisory and a few commercial databases offer designated remediation sections, and not all entries provide complete remediation information. Since \rs are often optional, they} \rev{are frequently absent from most databases, limiting their accessibility and practical application in security management. This absence poses significant threats to the security of the OSS community, as disclosing vulnerabilities without offering \rs creates potential attack windows.}
}

\begin{table}
\setlength{\tabcolsep}{3pt}
\centering
\scriptsize
\caption{Comparison of \RS Provided by Vulnerability DBs}
\scalebox{1}{
\begin{tabular}{@{}lllll|lllll@{}}
\toprule
\rowcolor[HTML]{EFEFEF} 
\textbf{DB name} & \textbf{Suggestion} & \textbf{Workaround} & \textbf{Desc} & \textbf{Ref link} & \cellcolor[HTML]{EFEFEF}\textbf{DB name} & \cellcolor[HTML]{EFEFEF}\textbf{Suggestion} & \cellcolor[HTML]{EFEFEF}\textbf{Workaround} & \cellcolor[HTML]{EFEFEF}\textbf{Desc} & \cellcolor[HTML]{EFEFEF}\textbf{Ref link} \\ \midrule
NVD~\cite{nvd}              & \emptycirc                   & \emptycirc                   & \halfcirc            & \halfcirc           & ExploitDB~\cite{edb}                                & \emptycirc                                           & \emptycirc                                           & \emptycirc                                     & \emptycirc                                    \\
\rowcolor[HTML]{EFEFEF} 
GitHubAdv~\cite{githubadvisory}        & \emptycirc                   & \halfcirc                  & \halfcirc            & \emptycirc            & \cellcolor[HTML]{EFEFEF}*SnykVulDB~\cite{snykvuldb}       & \cellcolor[HTML]{EFEFEF}\halfcirc                  & \cellcolor[HTML]{EFEFEF}\halfcirc                  & \cellcolor[HTML]{EFEFEF}\halfcirc            & \cellcolor[HTML]{EFEFEF}\halfcirc           \\
OSV~\cite{OSV}              & \emptycirc                   & \emptycirc                   & \halfcirc            & \emptycirc            & *Mend~\cite{mend}                                    & \halfcirc                                          & \emptycirc                                           & \halfcirc                                    & \emptycirc                                    \\
\rowcolor[HTML]{EFEFEF} 
SecList~\cite{seclists}          & \emptycirc                   & \emptycirc                   & \halfcirc            & \emptycirc            & \cellcolor[HTML]{EFEFEF}*VulDB~\cite{vuldb}          & \cellcolor[HTML]{EFEFEF}\halfcirc                  & \cellcolor[HTML]{EFEFEF}\emptycirc                    & \cellcolor[HTML]{EFEFEF}\halfcirc            & \cellcolor[HTML]{EFEFEF}\emptycirc             \\ \bottomrule
\end{tabular}
}
\begin{minipage}{\linewidth} 
\footnotesize
\emph{
    \\\textbf{Desc}: whether \rs is in the vulnerability description. \textbf{Ref link}: designated links for \rs. Commercial DBs are prefixed by *.
    }
\end{minipage}
\label{tab:vuldb}
\end{table}

\subsubsection{Study of \RS Categories on Vulnerability Databases}
\label{sec:advisory_rs}

To statistically understand the \rs provided by vulnerability DBs, we manually analyzed the descriptions for \rs for three non-commercial DBs. We fetched the CVE descriptions from these DBs and categorized the \rs according to our taxonomy. Due to the accessibility, we only focused on non-commercial DBs. Considering the diverse and abundant descriptions, NVD, GitHub Advisory, and OSV were included in our dataset. To prepare the data, we crawled the descriptions from DBs for the 3,044 CVEs linked to GitHub issues. As duplicates of descriptions exist, we first de-duplicated them and concatenated the different parts as \rev{a union}. 

Splitting the cases, the first three authors categorized the \rs into the taxonomy obtained before as in Figure~\ref{fig:advisory}. 
The lower bar depicts the distribution of \rs major categories from vulnerability DBs, and the upper one is from the GitHub issue.
It is observed that 54.85\% of CVEs lack remediation suggestions, underscoring the need for additional sources of \rs, such as community-driven platforms like GitHub. Fortunately, $93.26\%$ of issues have actionable \rs for these CVEs, and $85.67\%$ of them were accepted. 
\rev{For categories that are rarely present in vulnerability databases, such as \patch and \bypass, or appear less frequently than in GitHub issues, like \va, \rs from GitHub issues can significantly enhance coverage by introducing additional remediation options for CVEs.
Thus, community-sourced \rs can effectively complement the gaps in public vulnerability databases. Collecting and formatting these \rs could help optimize future remediation solutions, improving the overall effectiveness of vulnerability management.}

\mybox{
\rev{\textbf{Finding 9}: Although $54\%$ of CVEs lack remediation suggestions, highlighting the need for additional sources of \rs, $93.43\%$ of issues provide actionable \rs for these CVEs, with $85.67\%$ of them being accepted. 
Notably, the proportion of \patch could be boosted for 7.7x, while \bypass could be 2.0x. This suggests that \rs from community-driven platforms, such as GitHub issues, can effectively bridge the gaps left by public vulnerability databases, offering a promising direction for optimizing future remediation tools.}}




\section{Discussion}

\subsection{Threats of Validity}
\textbf{Internal Threat:} The main threat to our study is the use of manual labeling and categorization, which can inevitably introduce biases, including in \rs categorization, acceptance determination, and identification of issue status as well as summarization of reasons for rejections and re-openings. To counteract these biases and maintain a balance between efficiency and accuracy, we implemented a Hybrid Card Sorting supplemented by Cross-Review to compensate for the inaccuracy of labeling. 

\rev{Another threat is the inclusion of fixes for potentially non-security issues introduced by fuzzy GitHub searching. As our aim is to collect as many security issues as possible to analyze their \rs, GitHub issues via its keyword-based advanced search can provide a more comprehensive scope, even though it is inevitable to include issues not closely aligned with our main objective. To evaluate the relevance, we conducted a manual validation in Section~\ref{sec:data}, revealing the majority of the searched results were closely related to security and vulnerabilities. As the resolution of non-security issues can be beneficial to \rs taxonomy as well, we included their tactics in the \rs categories to honor the reproducibility and integrity of searched issues.}

\rev{A critical threat to our study is that our statistics and conclusions may not consistently generalize across the diverse contexts of all repositories. Given the considerable effort required, it is unrealistic to validate our findings in every possible context, thus we expanded our dataset as much as possible to obtain relatively representative results. Additionally, we conducted generalizability studies in Sections~\ref{sec:generalization} and~\ref{sec:lowstar}, which demonstrated that the adoption distribution of \rs remained generally consistent across major factors. Deviations and discrepancies were thoroughly analyzed and discussed in these sections.}

The last threat lies in the sample scrutinization for Vulnerability DBs. Due to the limited access to DBs and implicit \rs statements, a manual examination had to be conducted. The limited size of samples inevitably results in an incomplete overview of \rs. We have involved entries published in the last two decades to increase the diversity as much as possible. 

\noindent\textbf{External Threat:} The external threat to our study solely stems from the GitHub issue section, which may not capture the full scope of OSS security issue fixing, introducing a platform-specific bias. However, GitHub, as the largest source code hosting platform, offers extensive discussions on a wide array of software issues, permitting contributions from anyone, in contrast to project-oriented platforms like Apache Jira Board~\cite{apachejira} and Ubuntu Forum~\cite{ubuntuforum}, which are limited to specific products and topics. \rev{Additionally, GitHub provides free API access, facilitating smooth data crawling, unlike commercial platforms. While the GitHub issue may not represent entire OSS discussions, the vast volume of data and diverse range of topics provide significant analytical value.} 

\subsection{Implications}

\rev{The current challenges and potential solutions 
regarding OSS remediation are discussed.}

\noindent\textbf{\rev{Deficiency of RTs in Vulnerability DBs:}}
It is concerning that modern vulnerability DBs often lack \rs (missing from approximately 54\% of CVE entries), as these DBs are frequently referred to in SCA tool reports. The absence of remediation suggestions can significantly impede the security of downstream projects. Therefore, it is recommended that vulnerability DBs include a designated area for remediation suggestions. Additionally, these suggestions could be categorized into long-term solutions and short-term workarounds, offering users flexible remediation options suited to various circumstances, such as breaking changes in secure versions. Given that 
\rs are continuously evolving proven by 9,648 reopened issues, it is crucial for vulnerability DBs to regularly update and incorporate new solutions. The inclusion of comprehensive \rs can substantially support secure practices across all other four roles, thereby enhancing the overall security of the OSS ecosystem. 

\noindent\textbf{\rev{Insufficient Support of RTs in Remediation Tools:}}
Current SOTA remediation tools often lack a diverse range of suggestions, typically concentrating on upgrades, as highlighted in Related Work~\ref{sec: tool study}. However, considering the variety of \rs discussed within the community, these tools could benefit from incorporating a broader spectrum of suggestions. Furthermore, with the vast amount of remediation suggestions available, future tools could leverage advanced technologies such as Retrieval Augmented Generation, which can enable the automation of customized solutions, learning, and synthesizing tactics from extensive datasets to enhance the efficacy and applicability of remediation tools. Finally, given the evolving \rs, remediation tools should also timely update their \rs to accommodate the latest changes.

\noindent\textbf{\rev{More Actions Required for OSS Practitioners:}}
Both users and maintainers are concerned about the secondary problems brought by remediation, which potentially disrupt the software system if not properly executed. Ensuring that existing functionalities remain intact typically requires thorough testing and review. Governors in the OSS ecosystem and owners of OSS infrastructures should promote 
the adoption of remediation measures. Furthermore, the effectiveness of \rs may diminish over time. Insights from reopened issues indicate that those not completely resolved often lead to more effective solutions upon re-evaluation. Thus long-term monitoring and continuous optimization based on feedback are essential for the sustainable development of OSS security.


\noindent\textbf{\rev{Proof-of-concept to Promote \rs Adoption:}}
As revealed in Section~\ref{sec:rq2}, users generally favor simple yet effective remediation measures that can be quickly implemented with less concerns about the long-term effects. 
Therefore, automated or simplified solutions are more readily accepted by users, facilitating smoother and more efficient remediation processes for end-users.
Maintainers also prefer simplicity in remediation measures, but they prioritize stability and avoid increasing maintenance efforts. They usually require concrete proof and a clear incentive to adopt \rs, demonstrating the severity of current issues may motivate maintainers by emphasizing the threats they pose. Additionally, evidence of remediation effectiveness, similar to validation in Automatic Program Repair~\cite{tian2023best}, can justify the necessity of remediation and encourage broader adoption.

\noindent\rev{\textbf{Integration of \rs into Development Cycle:}
Given the fast-paced nature of modern software development, 
it is urgently required solutions that can help developers quickly and effortlessly validate whether RTs are successfully adopted without secondary issues. Continuous Integration/Continuous Deployment (CI/CD), served as the mainstream solution of agile development, can effectively
streamline the adoption of secure fixes. By automatically detecting security issues and recommending \rs within the development workflow, CI/CD systems can ensure that fixes are promptly applied while minimizing disruption. Additionally, automated testing and validation mechanisms can assess the impact of \rs on software stability and performance, increasing maintainers' confidence in adopting community-driven fixes.
}

\noindent\rev{\textbf{Encouraging Standardization of \rs:}
The lack of consistency in \rs across OSS projects makes it difficult for users to adopt security fixes efficiently. Establishing standardized \rs formats, similar to SBOM standards, could improve automation and usability in security workflows. By defining clear categories of RT with metadata, vulnerability DBs and security tools can better integrate remediation guidance. Moreover, standardization efforts could benefit from community incentives, such as fix bounties, to encourage participation in refining and expanding the RT taxonomy, ensuring its relevance and adoption across OSS ecosystems.
}

\section{Conclusion}
Our study on OSS security remediation has filled significant research gaps by analyzing 21,187 GitHub issues, developing a hierarchical taxonomy of \rs, and revealing that a significant portion of community-driven \rs remains unsupported by current tools (unseen in 44\% of studied issues). Key insights include that \rs should be tailored for different roles, such as preference on simple measures of downstream users and inclination on stability by maintainers. By analyzing the rejection reasons, the requirements of a competitive \rs are elaborated. 
Finally, the absent \rs for 54\% highlights the need for ongoing enhancements and support for \rs by remediation tools and vulnerability DBs to address the evolving landscape of OSS vulnerabilities. Future efforts should focus on broadening the scope of supported remediation tactics in tools and databases, ensuring continuous adaptation to emerging security challenges.

\section*{Acknowledgment}
This research is supported by the Ministry of Education, Singapore, under its Academic Research Fund Tier 1 (RG96/23). It is also supported by the National Research Foundation, Singapore, and DSO National Laboratories under the AI Singapore Programme (AISG Award No: AISG2-GC-2023-008); by the National Research Foundation Singapore and the Cyber Security Agency under the National Cybersecurity R\&D Programme (NCRP25-P04-TAICeN); and by the Prime Minister’s Office, Singapore under the Campus for Research Excellence and Technological Enterprise (CREATE) programme.
Any opinions, findings and conclusions, or recommendations expressed in these materials are those of the author(s) and do not reflect the views of National Research Foundation, Singapore, Cyber Security Agency of Singapore, Singapore.

\section*{Data Availability}
Our data and source code are available on our repository~\cite{dataset}.
\clearpage

\bibliographystyle{ACM-Reference-Format}
\bibliography{acmart}


\end{document}